\documentclass[english,aip,jcp,amsmath,amssymb,reprint]{revtex4-2}
\usepackage[T1]{fontenc}
\usepackage[utf8]{inputenc}
\usepackage{color}
\usepackage{babel}
\usepackage{textcomp}
\usepackage{amsmath}
\usepackage{amssymb}
\usepackage{graphicx}
\usepackage{hyperref}
\usepackage{dcolumn}

\makeatletter
\newcommand{\lyxdot}{.}

\DeclareUnicodeCharacter{2212}{-}
\makeatother
\preprint{AIP/123-QED}
\begin{document}
\title{Polariton assisted incoherent to coherent excitation
energy transfer between colloidal nanocrystal quantum dots}
\author{Kaiyue Peng}
\email{kaiyue_peng@berkeley.edu}
\affiliation{ 
Department of Chemistry, University of California,
Berkeley, California 94720, United States}
\author{Eran Rabani}
\email{eran.rabani@berkeley.edu}
\affiliation{ 
Department of Chemistry, University of California,
Berkeley, California 94720, United States}
\affiliation{Materials Sciences Division, Lawrence Berkeley National Laboratory,
Berkeley, California 94720, United States}
\affiliation{The Sackler Center for Computational Molecular and Materials Science, Tel Aviv University, Tel Aviv 69978, Israel}

\date{\today}

\begin{abstract}
We explore the dynamics of energy transfer between
two nanocrystal quantum dots placed within an optical microcavity.
By adjusting the coupling strength between the cavity photon mode
and the quantum dots, we have the capacity to fine-tune the effective
coupling between the donor and acceptor. Introducing a non-adiabatic
parameter, $\gamma$, governed by the coupling to the cavity mode,
we demonstrate the system's capability to shift from the overdamped
Förster regime ($\gamma\ll1$) to an underdamped coherent regime ($\gamma\gg1$).
In the latter regime, characterized by swift energy transfer rates,
the dynamics are influenced by decoherence times. To illustrate this,
we study the exciton energy transfer dynamics between two closely
positioned CdSe/CdS core/shell quantum dots with sizes and separations
relevant to experimental conditions. Employing an atomistic approach,
we calculate the excitonic level arrangement, exciton--phonon interactions,
and transition dipole moments of the quantum dots within the microcavity.
These parameters are then utilized to define a model Hamiltonian.
Subsequently, we apply a generalized non-Markovian quantum Redfield
equation to delineate the dynamics within the polaritonic framework.
\end{abstract}

\keywords{Quantum dots, Polariton, Micro--cavity, Coherent energy transfer}
\maketitle

\section{Introduction}
In the realm of nanophotonics and quantum optics, understanding and
controlling the transfer of energy between quantum emitters is paramount
for advancing technologies ranging from solar cells to quantum information
processing.\citep{choi2012layer,almutlaq2024engineering}
Colloidal nanocrystal (NC) quantum dots (QDs), with unique electronic
and optical properties,\citep{alivisatos1996perspectives,efros2000electronic,gomez2006optical}
have emerged as versatile platforms for studying energy transfer processes
at the nanoscale.\citep{klimov2007spectral,choi2012layer,kodaimati2018energy,roy2020electrostatically}
Among these processes, Fluorescence Resonance Energy Transfer (FRET)
has garnered considerable attention for its role in mediating efficient
energy transfer between quantum dots\citep{choi2012layer,kodaimati2017distance,roy2020electrostatically}
and other emitters,\citep{scholes2001adapting,barth2022unraveling}
and has been extensively studied and exploited in various fields,
including molecular biology,\citep{guo2018conformational,jares2003fret}
where it serves as a fundamental tool for probing molecular interactions
and dynamics.\citep{barth2022unraveling,kim2015real}
FRET, an incoherent energy transfer process,\citep{sun2011fret,zhong2023efficient}
relies on the weak coupling between a donor and an acceptor fluorophore,
with a transfer rate given by Fermi's golden rule (FGR):\citep{baer2008theory}
\begin{equation}
k_{\text{FGR}}=\frac{2\pi}{\hbar}\sum_{{\cal D}{\cal A}}\rho_{{\cal D}}\left|J_{{\cal DA}}\right|^{2}\delta\left(E_{{\cal D}}-E_{{\cal A}}\right).\label{eq:FGR rate}
\end{equation}
Here, $\rho_{{\cal D}}$ is the population distribution of excited
states on the donor, $J_{{\cal DA}}$ is the electromagnetic coupling
between transition densities connecting the ground state and excited
state for the donor and the acceptor (approximated by the dipole-dipole
term, see Eq.~(\ref{eq:dipole interatcion}) below), and $E_{{\cal D}}$
and $E_{{\cal A}}$ are the donor and acceptor excited state energies,
respectively. When the emission spectrum of the donor overlaps with
the absorption spectrum of the acceptor and they are within a certain
distance, typically within a few nanometers, energy transfer can occur with
high efficiency, but on relatively long timescales (a few tens of
nanoseconds).\citep{scholes2001adapting, kodaimati2017distance, loiudice2020tunable}
FRET processes are inherently limited by the steep distance dependence
($R^{-3}$) of the dipole-dipole coupling between the donor and acceptor
QDs, resulting in small couplings, overdamped energy transfer dynamics,
and long transfer rates, necessitating the exploration of alternative
mechanisms to accelerate and control the energy transfer mechanism.\citep{govorov2007theory,schachenmayer2015cavity,wang2021polariton} 

One strategy for regulating the exciton energy transfer process involves utilizing a donor--bridge--acceptor configuration, wherein the transition mechanism and rates can be adjusted by manipulating the states of the bridge.\citep{chowdhury2022interference, albinsson2010excitation} Recent advancements have led to the emergence of polariton--mediated energy transfer as a promising alternative to FRET.\citep{xu2023ultrafast,schachenmayer2015cavity,feist2015extraordinary,zhong2017energy} Polaritons, arising from the strong coupling between photons and excitons,\citep{weight2023investigating} present distinct advantages for facilitating energy transfer, including heightened coupling strength, tunable spectral overlap, and minimal dependence on distance, making efficient remote energy transfer possible with more flexible material design.\citep{du2018theory,zhong2017energy,kimura2009general,wang2021polariton,saez2018organic} As efforts intensify to exploit polariton-mediated energy transfer, ongoing research endeavors are dedicated to elucidating the control parameters dictating the efficiency and dynamics of these processes, such as energy decay rate, dephasing rate, and Rabi splitting/light coupling strength, as well as their scaling with number of molecules involved.\citep{reitz2018energy,coles2014polariton,wang2021polariton,xiang2020intermolecular}

\begin{figure}[t]
\begin{centering}
\includegraphics[width=8cm]{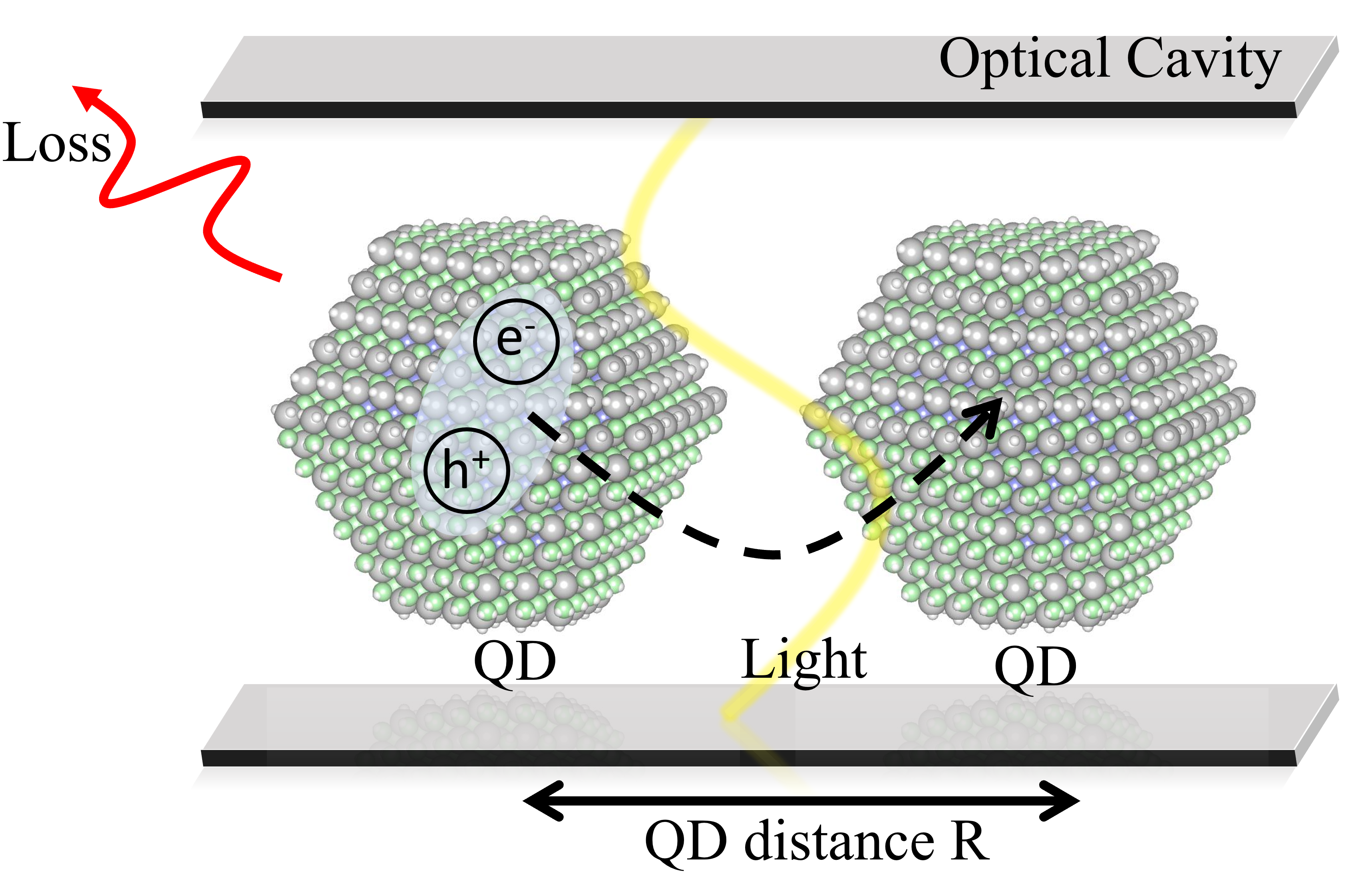}
\par\end{centering}
\caption{An illustration of two core/shell quantum dots in an optical microcavity.
\label{fig:illustration}}
\end{figure}

In this study, we examine the dynamics of energy transfer between
two quantum dots (QDs) situated in an optical microcavity, as depicted
in Fig.~\ref{fig:illustration}. By adjusting the coupling strength
between the cavity photon mode and the QDs, we can manipulate the
effective coupling between the donor and acceptor while maintaining
the system reorganization energy almost constant. We demonstrate that
this ability to control the effective coupling between the donor and
acceptor, without altering other aspects of the system, offers a means
to transition the system from the overdamped Förster regime to an
underdamped coherent regime, where the transfer rate is influenced
by decoherence time.

To illustrate this transition to coherent energy transfer dynamics,
we examine two closely positioned CdSe/CdS core/shell QDs of sizes
and separations relevant to experimental conditions. We utilize an
atomistic theory to compute the excitonic level structure,\citep{philbin2018electron,jasrasaria2021interplay}
exciton--phonon couplings,\citep{jasrasaria2021interplay,jasrasaria2022simulations}
and transition dipole moments of the two QDs within a microcavity
setting. To capture the intricate interplay among photons, phonons,
and excitons, we employ a generalized non-Markovian quantum Redfield
equation, which provides a framework for accurately describing both
populations and coherences under weak exciton--phonon coupling conditions,
suitable for CdSe and CdS materials. We employ a non-adiabatic parameter,
$\gamma$, influenced by the coupling to the cavity mode (among other
factors) and demonstrate that in the nonadiabatic regime ($\gamma\ll1$),
our approach aligns with Förster theory without the necessity of specifying
the spectral line width, as it is self-determined by the Redfield
approach. As $\gamma$ increases beyond $1$ towards the adiabatic
limit, the dynamics exhibit coherent oscillations at the Rabi frequency,
with energy transfer rates dictated by the decoherence time.

\section{Model and Dynamics\label{sec:Model-Hamiltonian}}
\subsection{Donor-acceptor and cavity Hamiltonians}
We write the total Hamiltonian, $H=H_{{\rm \mathcal{DA}}}+H_{{\rm cav}}$,
as a sum of the donor--acceptor vibronic Hamiltonian, $H_{{\rm \mathcal{DA}}}$,
and the contributions due to the cavity mode, $H_{{\rm cav}}$. The
donor--acceptor vibronic Hamiltonian of the two QDs is derived from
the model Hamiltonian developed by Jasrasaria and Rabani.\citep{jasrasaria2022simulations,jasrasaria2021interplay}
This model describes hot exciton relaxation in a single QD, where
ultrafast relaxation dynamics are driven by multi--phonon channels.\citep{jasrasaria2023circumventing}
Its generalization for a donor--acceptor system is given by
\begin{align}
H_{{\rm \mathcal{DA}}} & =\sum_{n\in\mathcal{D},\mathcal{A}}E_{n}\left|\psi_{n}\right\rangle \left\langle \psi_{n}\right|\nonumber \\ &+\sum_{n\in\mathcal{D},m\in\mathcal{A}}J_{nm}\left(\left|\psi_{n}\right\rangle \left\langle \psi_{m}\right|+h.c.\right)\nonumber \\
 & +\sum_{\alpha}\hbar\omega_{\alpha}b_{\alpha}^{\dagger}b_{\alpha}+\sum_{\alpha}\sum_{n,m\in\mathcal{D},\mathcal{A}}V_{nm}^{\alpha}\left|\psi_{n}\right\rangle \left\langle \psi_{m}\right|q_{\alpha}.\label{eq:FRET_H}
\end{align}
The first and second terms on the right hand side describe the manifold of donor and acceptor excitonic states and their couplings, respectively,
and the remaining terms describe the lattice vibrations and the exciton--phonon
couplings, approximated to lowest order in the phonons displacement,
$q_{\alpha}$. In the above, $E_{n}$ is the energy of a singly excited
state $\left|\psi_{n}\right\rangle$, with $n\in\mathcal{D}/\mathcal{A}$
and $\mathcal{D}=\left\{ {\cal D}_{1},{\cal D}_{2},\cdots\right\} $,
$\mathcal{A}=\left\{ {\cal A}_{1},{\cal A}_{2},\cdots\right\} $.
We use the notation $\left|\psi_{{\cal D}_{1}}\right\rangle $ to
describe the donor in excitonic state ${\cal D}_{1}$ while the acceptor
is in its ground state and vise versa for $\left|\psi_{{\cal A}_{1}}\right\rangle $.
The excitonic energies and wave functions ($E_{n}$ and $\left|\psi_{n}\right\rangle $)
were obtained by ignoring the coupling between the donor and acceptor,
using the semi--empirical pseudopotential methods combined with the
Bethe--Salpeter equation (BSE).\citep{jasrasaria2022simulations}
The quasiparticle states near the top of the valence band and the
bottom of the conduction band were generated for each QD using the
filter diagonalization technique\citep{wall1995extraction,toledo2002very}
with a real--space grid basis of $N_{g}\approx2\cdot10^{6}$ and
a grid spacing of $\approx0.75 a_{0}$. A total of $80$ electron (unoccupied)
and $140$ hole (occupied) states were generated to construct the
Bethe--Salpeter Hamiltonian,\citep{rohlfing2000electron}
within the static screening approximation,\citep{jasrasaria2022simulations}
providing converged excitonic energies and transition dipole moments
for the lowest $30$ excitonic states that participate in the dynamics,
for each QD.

The coupling between the donor in state $\left|\psi_{\mathcal{D}}\right\rangle $ and the acceptor in state $\left|\psi_{\mathcal{A}}\right\rangle $, $J_{\mathcal{DA}}$, was approximated to lowest order in the corresponding transition charge densities, with the dipole--dipole leading term given by:\citep{baer2008theory} 
\begin{equation}
J_{\mathcal{DA}}=f_{{\cal D}}f_{{\cal A}}\frac{\left|\boldsymbol{\mu}_{{\cal D}}^{{\rm ex}}\right|\left|\boldsymbol{\mu}_{{\cal A}}^{{\rm ex}}\right|}{\epsilon_{\text{m}}\left|R\right|^{3}}\kappa\left(\theta_{{\cal DA}},\theta_{{\cal D}},\theta_{{\cal A}}\right)\label{eq:dipole interatcion}
\end{equation}
Here, $\boldsymbol{\mu}_{\mathcal{D\text{/}A}}^{{\rm ex}}=\left\langle \psi_{\mathcal{D\text{/}A}}\right|\hat{\boldsymbol{\mu}}\left|\psi_{g}\right\rangle $
is the transition dipole moment obtained from the pseudopotential
model~\citep{jasrasaria2022simulations} and $\left|\psi_{g}\right\rangle $
is the ground state. The local field factor in the above is given
by $f_{\mathcal{D}/\mathcal{A}}=\frac{3}{2+\epsilon_{\text{\ensuremath{\mathcal{D}}/\ensuremath{\mathcal{A}}}}/\epsilon_{\text{m}}}=\frac{3}{7}$
for~\citep{wang1996pseudopotential} $\epsilon_{{\cal D}}=\epsilon_{{\cal A}}=5$ and $\epsilon{}_{\text{m}}=1$,
and $\left|\kappa\left(\theta_{{\cal DA}},\theta_{{\cal D}},\theta_{{\cal A}}\right)\right|\le2$
is an angle dependent term which can be approximated by averaging
over the angles, with the well-known results that $\sqrt{\left\langle \kappa^{2}\right\rangle }=\sqrt{\frac{2}{3}}$.\citep{latt1965energy}
Alternatively, one can compute the angles $\theta_{{\cal DA}}$, $\theta_{{\cal D}}$,
and $\theta_{{\cal A}}$, which depend on the relative orientation
of the transition dipole moments.\citep{fedchenia1994influence} 

The donor and acceptor excitonic states also couple to the lattice
vibrations with frequencies $\omega_{\alpha}$ and modes $q_{\alpha}=\sqrt{\frac{\hbar}{2\omega_{\alpha}}}\left(b_{\alpha}^{\text{\ensuremath{\dagger}}}+b_{\alpha}\right)$,
obtained by diagonalizing the dynamical matrix of the QD donor and acceptor using the Stilling--Weber force field parameterized for Cd, Se, S elements.\citep{zhou2013stillinger} The exciton--phonon couplings were expanded to lowest order in the displacements, $q_{\alpha}$ and the strength of coupling between exciton $\left|\psi_{n}\right\rangle$ and $\left|\psi_{m}\right\rangle$ to mode $\alpha$, $V_{nm}^{\alpha}$, was obtained from the pseudopotential Hamiltonian.\citep{jasrasaria2021interplay}  We considered both diagonal $V_{n=m}^{\alpha}$ and off-diagonal $V_{n\ne m}^{\alpha}$ coupling terms and assumed that the excitons couple to the lattice modes on either the donor or the acceptor, depending on whether they localize to the donor or the acceptor.
The coupling of excitons to the lattice vibrations is key in accounting for the reorganization and the relaxation of the system to thermal equilibrium. 

The optical cavity was modeled by a single--mode Pauli--Fierz Hamiltonian within the dipole approximation and rotation wave approximation:\citep{mandal2020polariton,peng2023polaritonic}
\begin{align}
H_{\text{cav}} & =\hbar\omega_{c}a^{\dagger}a+\sum_{n\in\mathcal{D},\mathcal{A}}\hbar g_{n}\left(a^{\dagger}\left|\psi_{g}\right\rangle \left\langle \psi_{n}\right|+h.c.\right)\nonumber \\
 & +\frac{\hbar}{\omega_{c}}\sum_{n\in\mathcal{D},\mathcal{A}}g_{n}^{2}\left|\psi_{g}\right\rangle \left\langle \psi_{g}\right|\nonumber \\
 & +\frac{\hbar}{\omega_{c}}\sum_{n,m\in\mathcal{D},\mathcal{A}}g_{n}g_{m}\left(\left|\psi_{n}\right\rangle \left\langle \psi_{m}\right|+h.c.\right),\label{eq:Hamil_cav}
\end{align}
where $a$ ($a^{\dagger}$) is the annihilation (creation) operator
of a photon with frequency $\omega_{c}$, and $\hbar g_{n}$ is coupling strength between the cavity mode and an excitonic state $\left|\psi_{n}\right\rangle$ ($n\in\mathcal{D}/\mathcal{A}$). The cavity--exciton coupling strength is given by $g_{n}=\sqrt{\omega_{c}/\left(2\hbar\epsilon_{\text{m}}\mathcal{V}\right)}\boldsymbol{\mu}_{n}^{{\rm ex}}\cdot\boldsymbol{\varepsilon}$, where as before, $\boldsymbol{\mu}_{n}^{{\rm ex}}$ is the transition dipole moment from the ground state $\left|\psi_{g}\right\rangle$ to exciton $\left|\psi_{n}\right\rangle $ (obtained from the BSE calculation) and $\boldsymbol{\varepsilon}$ is a unit vector of the polarization direction of the electromagnetic field. $\epsilon_{{\rm m}}$ is the effective permittivity inside the cavity (medium) and $\mathcal{V}$ is the effective cavity quantization volume. The last two terms in the above equation correspond to the quadratic dipole self-energy, which has shown to be important in the ultra-strong coupling limit.\cite{Rokaj2018} The dipole self-energy term arises from the Power-Zienau-Woolley transformation that explicitly mixes the electronic and photonic degrees of freedom.\cite{sidler2023numerically} For the results reported in this work, the effect of the dipole-self-energy is negligible since $\omega_{c}\gg g_{n}$.

\subsection{Polaritonic representation and polaron transformation}
The total Hamiltonian can also be expressed using the polaritonic
basis, $\left\{ \left|\varphi_{n}\right\rangle =c_{ng}\left|\psi_{g},1\right\rangle +\sum_{m}c_{nm}^{\text{ex}}\left|\psi_{m},0\right\rangle \right\} $, where $\left|\psi_{g},1\right\rangle $ and $\left|\psi_{m},0\right\rangle $ are the ground state with one photon or an excitonic state with zero photons, respectively. This representation simplifies the description of the dynamics within the Redfield approach described below, and is not limited by the weak donor-acceptor coupling strength (provided that the exciton-phonon couplings are small), as is the case for FRET between CdSe/CdS core-shell QDs. The coefficients $c_{ng}$ and $c_{nm}^{\text{ex}}$ were obtained by diagonalizing the system Hamiltonian (describing only the excitons and photons without the phonons), defined by:
\begin{align}
H_{{\rm S}} & =H_{\text{cav}}+\underset{n\in\mathcal{D},\mathcal{A}}{\sum}E_{n}\left|\psi_{n}\right\rangle \left\langle \psi_{n}\right|\nonumber \\
 & +\underset{n\in\mathcal{D},m\in\mathcal{A}}{\sum}J_{nm}\left(\left|\psi_{n}\right\rangle \left\langle \psi_{m}\right|+h.c.\right).
 \label{eq:Hs}
\end{align}
In the above, we assumed that the ground state energy is zero, namely $E_{g}=0$. Using the polaritonic basis obtained by diagonalizing the system Hamiltonian in Eq.~\eqref{eq:Hs}, the total Hamiltonian ($\mathcal{H}=\mathcal{H}_{{\rm S}}+\mathcal{H}_{\text{B}}+\mathcal{H}_{\text{I}}$) given by $H_{{\rm \mathcal{DA}}}+H_{{\rm cav}}$ can be written as:
\begin{align}
{\cal H} & =\overset{{\scriptstyle \mathcal{H}_{{\rm S}}}}{\overbrace{\sum_{n}\left(\tilde{E}_{n}-\lambda_{n}\right)\left|\varphi_{n}\right\rangle \left\langle \varphi_{n}\right|}}\nonumber \\
 & +\overset{{\scriptstyle \mathcal{H}_{\text{B}}}}{\overbrace{\sum_{\alpha}\hbar\omega_{\alpha}b_{\alpha}^{\text{\ensuremath{\dagger}}}b_{\alpha}}+}\overset{{\scriptstyle \mathcal{H}_{\text{I}}}}{\overbrace{\sum_{n\neq m}W_{nm}\left|\varphi_{n}\right\rangle \left\langle \varphi_{m}\right|}}.
 \label{eq:Hredfield}
\end{align}
In the above, $\tilde{E}_{n}$ is the polaritonic energy of state $\left|\varphi_{n}\right\rangle$ (an eigenvalue of $H_{{\rm S}}$ in Eq.~\eqref{eq:Hs}) and $\lambda_{n}=\frac{1}{2}\sum_{\alpha}\left(\tilde{V}_{nn}^{\alpha}\right)^{2}/\omega_{\alpha}^{2}$ is the corresponding reorganization energy (polaron shift). The dressed coupling between polaritons and phonons, $W_{nm}$, is given by:
\begin{align}
W_{nm} & =\exp\left(-\frac{i}{\hbar}\sum_{\gamma}\frac{p_{\gamma}\tilde{V}_{nn}^{\gamma}}{\omega_{\gamma}^{2}}\right)\times\nonumber \\
 & \times\left[\sum_{\alpha}\tilde{V}_{nm}^{\alpha}q_{\alpha}\right]\exp\left(+\frac{i}{\hbar}\sum_{\xi}\frac{p_{\xi}\tilde{V}_{mm}^{\xi}}{\omega_{\xi}^{2}}\right).
\end{align}
To derive the Hamiltonian in Eq.~\eqref{eq:Hredfield} we also performed a small polaron transformation (in addition to diagonalizing the system Hamiltonian to form the polaritonic states), as detailed in Refs.~\citenum{jasrasaria2023circumventing} and \citenum{peng2023polaritonic}. As a result of the polaron transformation, the dressed coupling between polaritons, $W_{nm}$, depends exponentially on the momenta ($p_{\alpha}$) and position ($q_{\alpha}$) of all phonon modes, enabling the description of multi–phonon relaxation channels within the lowest order perturbation in $\mathcal{H}_{\text{I}}$. In the above, the dressed coupling (and the reorganization energy) depend on $\tilde{V}_{nm}^{\alpha}$, the transformed coupling between two polaritonic states $\left|\varphi_{n}\right\rangle$  and $\left|\varphi_{m}\right\rangle$  and phonon mode $\alpha$, obtained by applying the unitary transformation $U$ that diagonalizes $H_{{\rm S}}$ to $\tilde{V}^{\alpha}=U^{\dagger}V^{\alpha}U$, for each mode $\alpha$.

\subsection{Dynamics}
In this work, we leverage on the fact that the rescaled coupling between polaritons and phonon, $\mathcal{H}_{\text{I}}$ (Eq.~\eqref{eq:Hredfield}), can be treated perturbatively. Thus, to generate the dynamics within the Hilbert space of polaritons, we adopt the Redfield equations instead of using more advanced methods suitable for strong coupling.\citep{perez2023simulating} The equation of motion for the reduced density matrix, $\sigma\left(t\right)$,
representing the polaritonic subspace can then be written as:\citep{nitzan2006chemical}\begin{widetext}
\begin{align}
\frac{d\sigma\left(t\right)}{dt} & =-\frac{i}{\hbar}\left[\mathcal{H}_{\text{S}}+\left\langle \mathcal{H}_{\text{I}}\right\rangle _{\text{B}},\sigma\left(t\right)\right]\nonumber \\
- & \frac{1}{\hbar^{2}}\sum_{nm,kl}\int_{0}^{t}d\tau\left\{ C_{nm,kl}\left(\tau\right)\left[\left(\mathcal{H}_{\text{S}}^{{\rm I}}\right)_{nm},e^{-i/\hbar\mathcal{H}_{\text{S}}\tau}\left(\mathcal{H}_{\text{S}}^{{\rm I}}\right)_{kl}\sigma\left(t-\tau\right)e^{i/\hbar\mathcal{H}_{\text{S}}\tau}\right]\right.\nonumber \\
- & \left.C_{nm,kl}^{*}\left(\tau\right)\left[\left(\mathcal{H}_{\text{S}}^{{\rm I}}\right)_{nm},e^{-i/\hbar\mathcal{H}_{\text{S}}\tau}\sigma\left(t-\tau\right)\left(\mathcal{H}_{\text{S}}^{{\rm I}}\right)_{kl}e^{i/\hbar\mathcal{H}_{\text{S}}\tau}\right]\right\}
+ \mathcal{L_{\rm loss}} \sigma (t)
,\label{eq:gqme}
\end{align}
\end{widetext}
where the polariton-phonon coupling is expressed as $\mathcal{H}_{\text{I}}=\sum_{nm}W_{nm}\left(\mathcal{H}_{\text{S}}^{{\rm I}}\right)_{nm}$
with $\left(\mathcal{H}_{\text{S}}^{{\rm I}}\right)_{nm}=\left(1-\delta_{nm}\right)\left|\varphi_{n}\right\rangle \left\langle \varphi_{m}\right|$.  In the above equation, $\left\langle \cdots\right\rangle _{\text{B}}$
denotes an equilibrium thermal average over the so called ''bath''
degrees of freedom. The bath correlation function appearing above is given by:
\begin{align}
C_{nm,kl}\left(t\right) & =\left\langle W_{nm}\left(t\right)W_{kl}\left(0\right)\right\rangle _{\text{B}}-\left\langle W_{nm}\right\rangle _{\text{B}}\left\langle W_{kl}\right\rangle _{\text{B}},
\end{align}
with $W_{nm}\left(t\right)=e^{i/\hbar\mathcal{H}_{\text{B}}t}W_{nm}e^{-i/\hbar\mathcal{H}_{\text{B}}t}$.
These thermal averages and thermal correlation functions were computed
analytically for the above Hamiltonian as further described in the
supporting information. 

In addition, to describe cavity losses, we added a phenomenological loss term represented by the last term on the right hand side of Eq.~\eqref{eq:gqme},  given by:
\begin{align}
\left[\mathcal{L_{\text{loss}}}\sigma\left(t\right)\right]_{nm}=-\delta_{nm}\left(\frac{c_{ng}^{2}}{\tau^{\text{cav}}}\right)\sigma_{nm}\left(t\right),
\end{align}
where $\sigma_{nm}\left(t\right)=\left\langle \varphi_{n}\left|\sigma\left(t\right)\right|\varphi_{m}\right\rangle$, and the cavity loss rate from polaritonic state $\left|\varphi_{n}\right\rangle$ is give by the ratio of the cavity fraction of the polaritonic state, $c_{ng}^{2}=\left|\left\langle \psi_{g},1|\varphi_{n}\right\rangle \right|^{2}$, and the cavity lifetime, $\tau^{\text{cav}}$.

\section{Exciton energy transfer between QDs\label{sec:Exciton-energy-transfer}}
\begin{figure}[t]
\includegraphics[width=8cm]{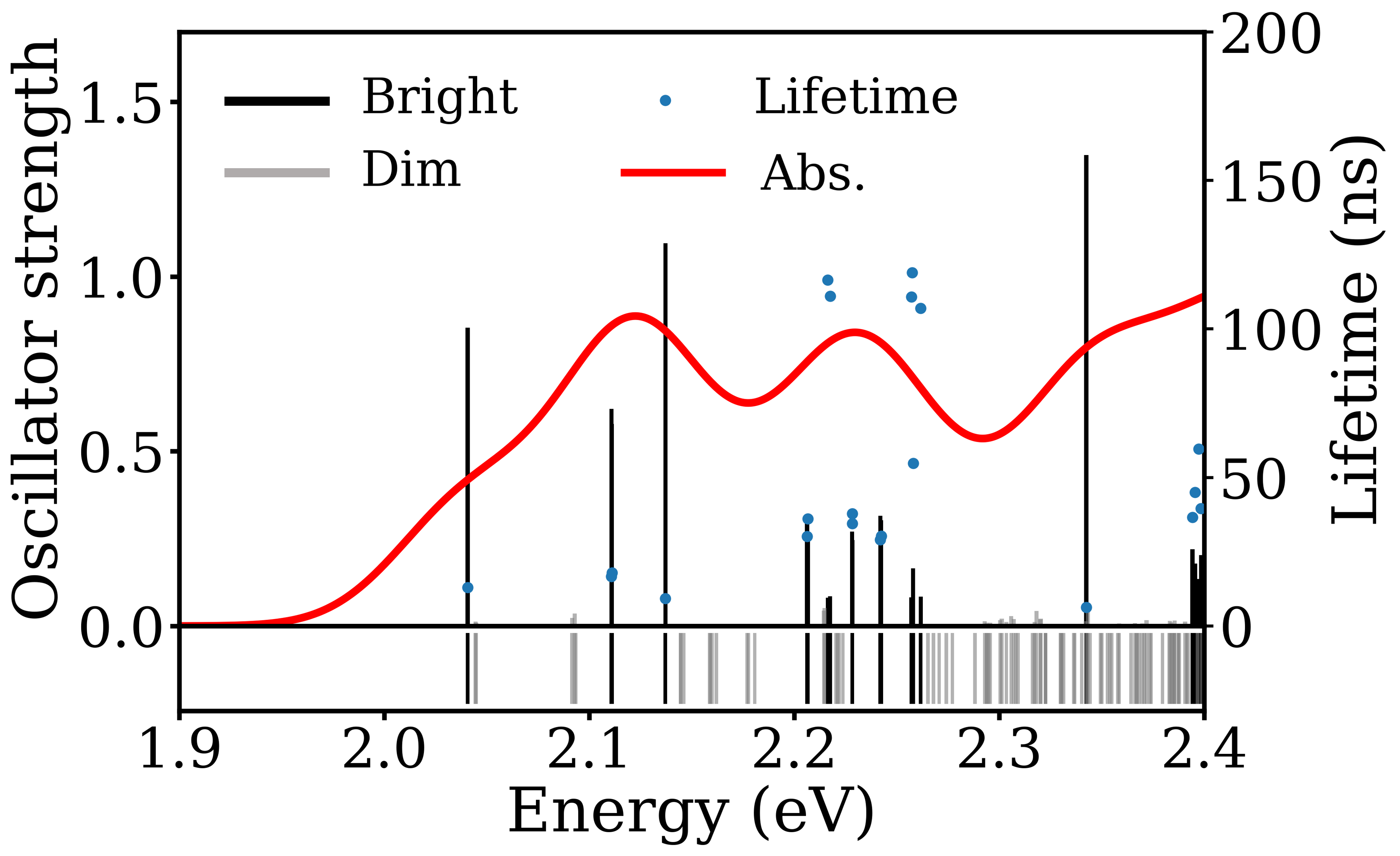}
\caption{The calculated linear absorption spectrum for CdSe/CdS core--shell
NCs, with a CdSe core diameter of $3$~nm and $2$ monolayers of
CdS shell. The vertical lines in upper subplot indicate the magnitude
of the oscillator strength of the transition from the ground state
to that excitonic state. The continuous absorption spectra (red curve)
was obtained by broadening the individual transitions with a variance
of $35\text{ meV}$ Gaussian function. The blue dots indicate the
radiative lifetime of each excitonic state given by Eq. (\ref{eq:radaitive_LT_ex})
below. The lower subplot in this figure shows the density of excitonic
states.\label{fig:spec} }
\end{figure}

In this section, we examine the excitonic structure of a typical $3$~nm
CdSe core with a 2-monolayer CdS shell and compare the predicted radiative
lifetimes to experimental data. Additionally, we demonstrate that
the Redfield approach, which treats the dressed exciton--phonon coupling
perturbatively but includes all orders of the hybridization $J_{nm}$,
aligns well with Förster theory in the absence of a cavity. The main
purpose of this section is to provide further validation for the approach
developed herein and in particular, for the calculated oscillator
strengths and vibronic couplings, essential to accurately describe
the energy transfer mechanisms and rates.

\begin{figure}[t]
\includegraphics[width=8cm]{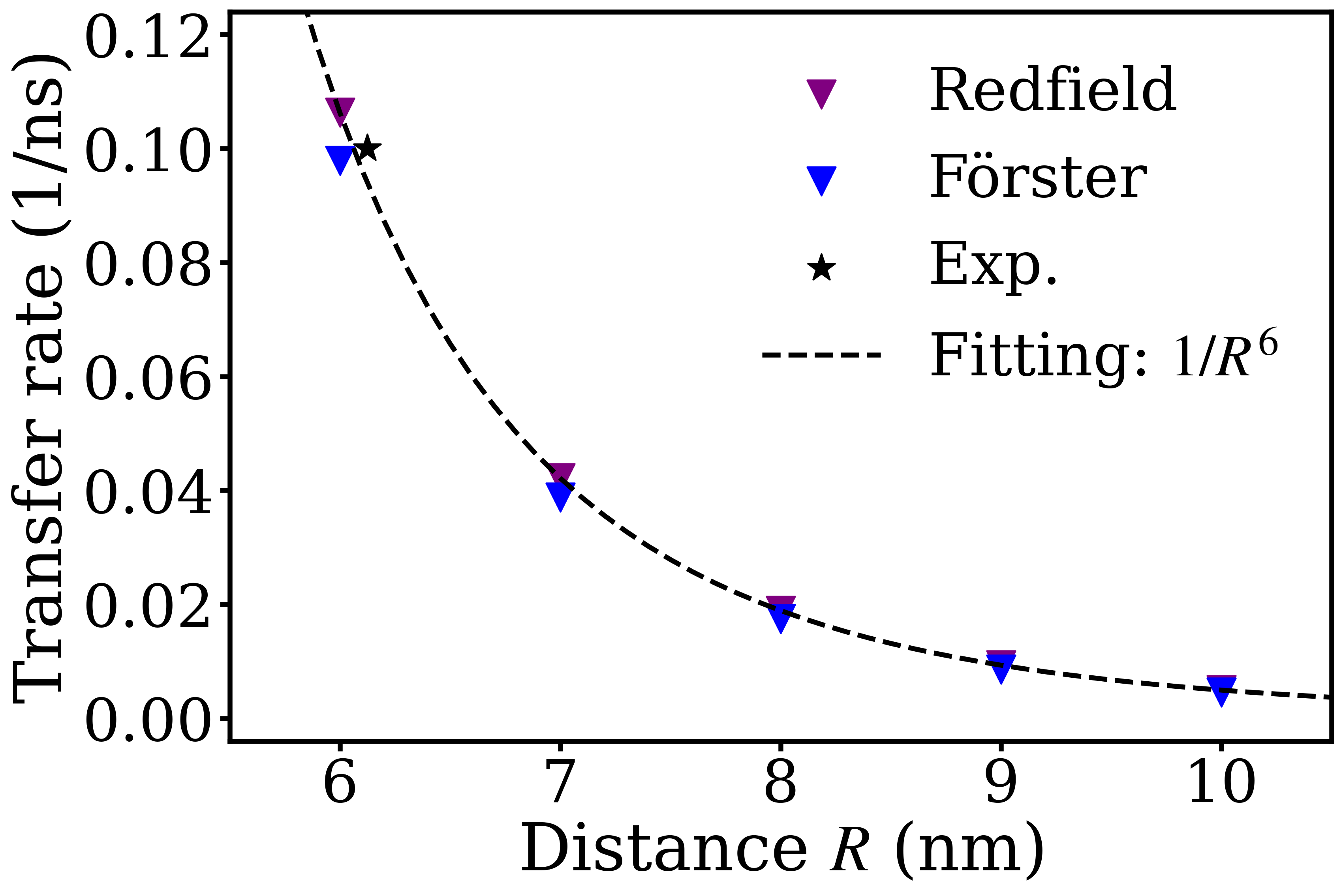}
\caption{Excitation energy transfer rates between two QDS, one with a 3.9nm CdSe core and 3 monolayers CdS shell, the other with a 3.9nm CdSe core 4 monolayers CdS shell, as a function of center-to-center distances between QDs, at 300K. The purple triangles represent the rates obtained from Eq. (\ref{eq:gqme}) (labeled as ``Redfield''); the blue triangles depict the rates calculated directly from Fermi's Golden Rule (Förster theory) as in Eq. (\ref{eq:FGR rate}) (labeled as ``Förster''). The black star is an experimental measurement between two CdSe QDs of different sizes, 3.85nm and 6.2nm in diameter, respectively.\citep{kagan1996electronic} The dash line depicts a $R^{-6}$ relationship and is a guide to the eye.\label{fig:FRET_rate_3.9}}
\end{figure}

\begin{figure*}[tp]
\includegraphics[width=15cm]{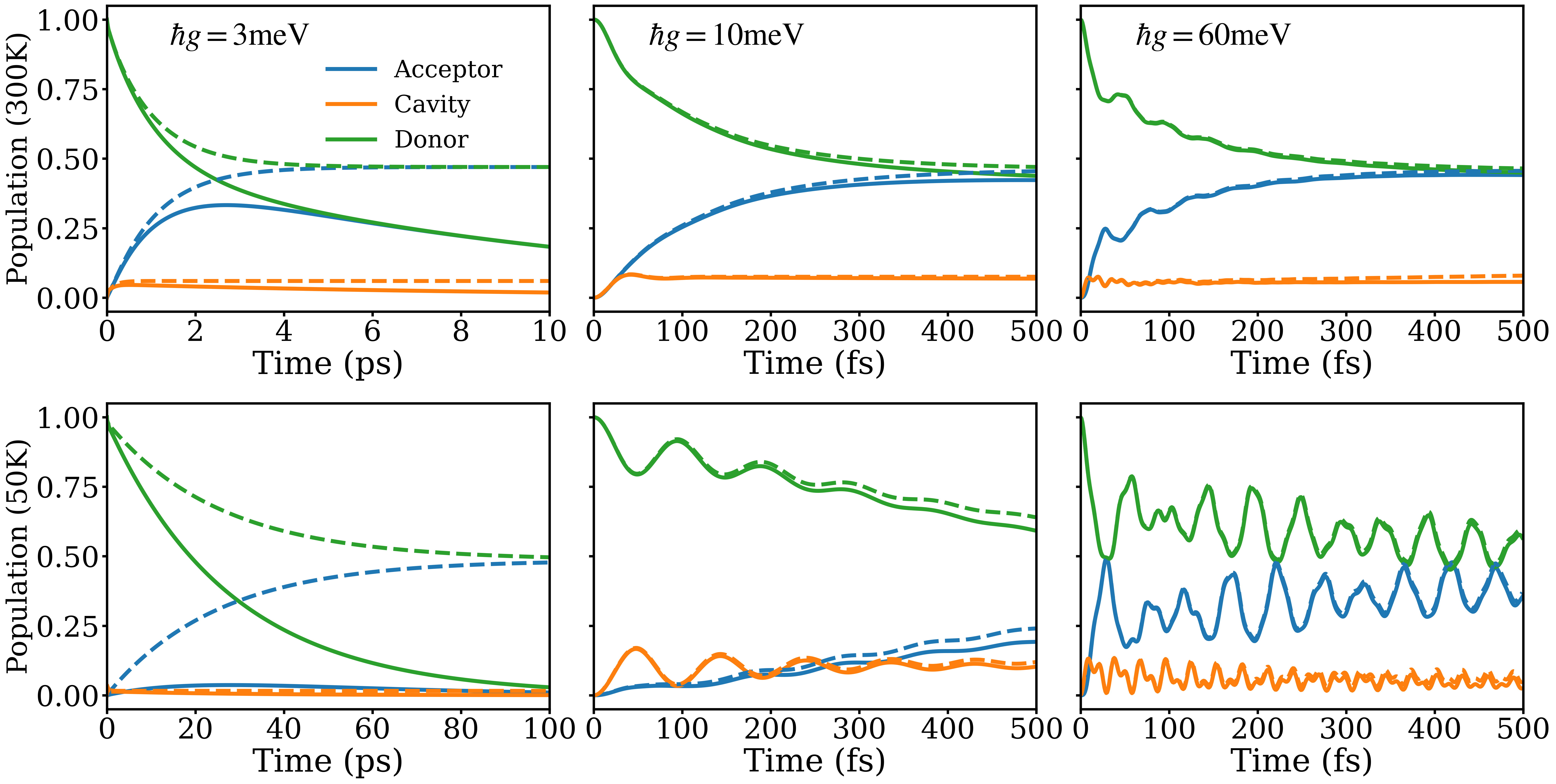}
\caption{Population dynamics for different light--matter interaction strengths
and different temperatures. The dashed and solid lines represent results
from an ideal and lossy cavity, respectively. The donor--acceptor
distance was fixed at $6\thinspace\text{nm}$. Three different QD-cavity
couplings were considered, $3\thinspace\text{meV}$, $10\,\text{meV}$,
and $60\,\text{meV}$, corresponding to effective coupling of $0.5\thinspace{\rm meV}$,
$4\thinspace{\rm meV}$, and $20\thinspace{\rm meV}$, respectively.\label{fig:Population-dynamics}}
\end{figure*}

Fig.~\ref{fig:spec} presents the calculated absorption spectrum
of a typical $3$~nm CdSe core with a 2-monolayer CdS shell, alongside
the density of excitonic states. The absorption onset is marked by
several distinct ``bright'' excitonic transitions, highlighted by
vertical lines representing the oscillator strength. As excitonic
energy increases, the density of excitons rises rapidly, with the
appearance of many ``dim'' transitions arising from the high density
of holes.\citep{brosseau2023ultrafast,jasrasaria2023circumventing}
The bright excitons play a crucial role in the energy transfer between
the donor and acceptor, whereas the dim excitons are essential to
account for dephasing and relaxation. To accurately describe both
energy transfer dynamics and the relaxation to equilibrium, we include
the $30$ lowest excitonic states for each QD, namely $30$ states
for the donor and the acceptor. Fig.~\ref{fig:spec} also shows the
calculated radiative lifetime for each excitonic state, obtained from
\begin{equation}
\tau_{n}^{{\rm ex}}=\frac{3c^{3}}{4E_{n}^{3}\left|\boldsymbol{\mu}_{n}^{{\rm ex}}\right|^{2}}.\label{eq:radaitive_LT_ex}
\end{equation}
The thermal average radiative lifetime, $\tau^{\text{ex}}=\sum_{n}P_{n}\tau_{n}^{\text{ex}}$
($P_{n}$ is the Boltzmann population of exciton $|\psi_{n}\rangle$),
is found to be $24\thinspace\text{ns}$, in very good agreement with
experimental measurements, ranging between $21-28$~ns.\citep{rabouw2015delayed,bae2013controlled,bae2013controlling,crooker2002spectrally} 

In Fig.~\ref{fig:FRET_rate_3.9}, we plot the calculated energy transfer rates as a function of the distance between the two QDs in the absence of a cavity. This is used to assess the accuracy of the Redfield approach in describing the energy transfer dynamics in the FRET limit, which can be compared to Förster perturbation theory as well as experimental values. The Redfield rates were obtained by fitting the calculated longtime population dynamics, $\rho_{{\cal D}/{\cal A}}\left(t\right)=\sum_{n\in{\cal {\cal D}}/{\cal A}}\sigma_{nn}\left(t\right)$, to an exponential decay:
\begin{equation}
\rho_{{\cal D}}\left(t\right)=\left(1-P_{{\cal D}}^{\text{eq}}\right)\exp\left[-k_{\text{ET}}t\right]+P_{{\cal D}}^{\text{eq}},\label{eq:Exp_fitting}
\end{equation}
where $P_{{\cal D}}^{\text{eq}}$ is the donor total exciton population
at equilibrium and $k_{\text{ET}}$ is the energy transition rate. For implementing Förster theory (Eq. (\ref{eq:FGR rate})), we have
convoluted the absorption transition with a Gaussian broadening function, with a width ($\approx60\thinspace{\rm meV}$) taken from the calculated photoluminescence spectrum.\citep{kailai2023lumi} The energy transfer rates calculated by both methods agree quite well with each other (note that the Förster theory rate depends on the choice of the broadening) and both recover the $R^{-6}$ distance dependence, as expected.  The calculated results are within a reasonable range compared to experimental measurements,\citep{kagan1996electronic,lingley2011high} also shown in Fig.~\ref{fig:FRET_rate_3.9}.

\section{Polaritonic energy transfer between QDs\label{sec:Polaritonic-energy-transfer}}
We now turn to discuss the role of the cavity mode
in the exciton transfer dynamics. We consider the transfer between
two CdSe/CdS core-shell QDs with $3$~nm core size and 2 layers of
CdS shell and a core-to-core distance of $6$~nm. The photon energy
was set to $\hbar\omega_{c}=2.06\thinspace\text{eV}$, slightly
above the absorption onset of the QDs and the orientation of the QDs
was fixed with the dipole moment parallel to the cavity polarization
vector. The cavity coupling strength to the lowest donor excited
state $\hbar g_{{\cal D}_{1}}\equiv\hbar g$ is used as a measure
of the QD-cavity coupling strength. We considered
two scenarios, an ideal cavity without energy loss and a more realistic
cavity with a photon escape time of\citep{le2018colloidal} $\tau_{\text{cav}}=0.5\,\text{ps}$.
We prepare the system in an initial thermal population of the excitonic
states on the donor and a canonical distribution of all phonon modes. 

Fig.~\ref{fig:Population-dynamics} shows representative population
dynamics for an ideal cavity (dashed lines) and a ``lossy'' cavity
(solid lines) at two different temperatures and varying cavity-QD
coupling values. The behavior of the donor and acceptor populations
is qualitatively similar for the ideal and lossy cavities on timescales
relevant to energy transfer. 
At low QD-cavity couplings ($\hbar g=3\text{\,meV}$),
the donor population dynamics exhibit a single exponential decay with
a rate approximated by Eq.~(\ref{eq:Exp_fitting}). The coupling
of both QDs to the same cavity mode results in an effective direct
coupling between them (see supporting information), estimated by tunneling
splitting $J_{{\rm eff}}=\frac{1}{2}\left(\tilde{E}_{{\rm LP+1}}-\tilde{E}_{{\rm LP}}\right)$,\citep{davis2009effect,skourtis2001electron,priyadarshy1996bridge}
where $\tilde{E}_{{\rm LP}}$ and $\tilde{E}_{{\rm LP+1}}$ are the two lowest polaritonic
energies obtained by diagonalizing $H_{{\rm S}}$, as described above.
This effective coupling is much larger than the direct dipole-dipole
coupling given by Eq.~(\ref{eq:dipole interatcion}). As $\hbar g$
increases above $60\thinspace\text{meV}$, the population dynamics
display pronounced oscillations with a frequency approximately equal
to $2J_{\text{eff}}$. In this case, the exciton transfer rate is
governed by decoherence time, especially at lower temperatures. A
similar transition from incoherent to coherent dynamics has recently
been reported for electron transfer between fused NC QDs.\citep{hou_incoherent_2023}

\begin{figure*}[t]
\includegraphics[width=15cm]{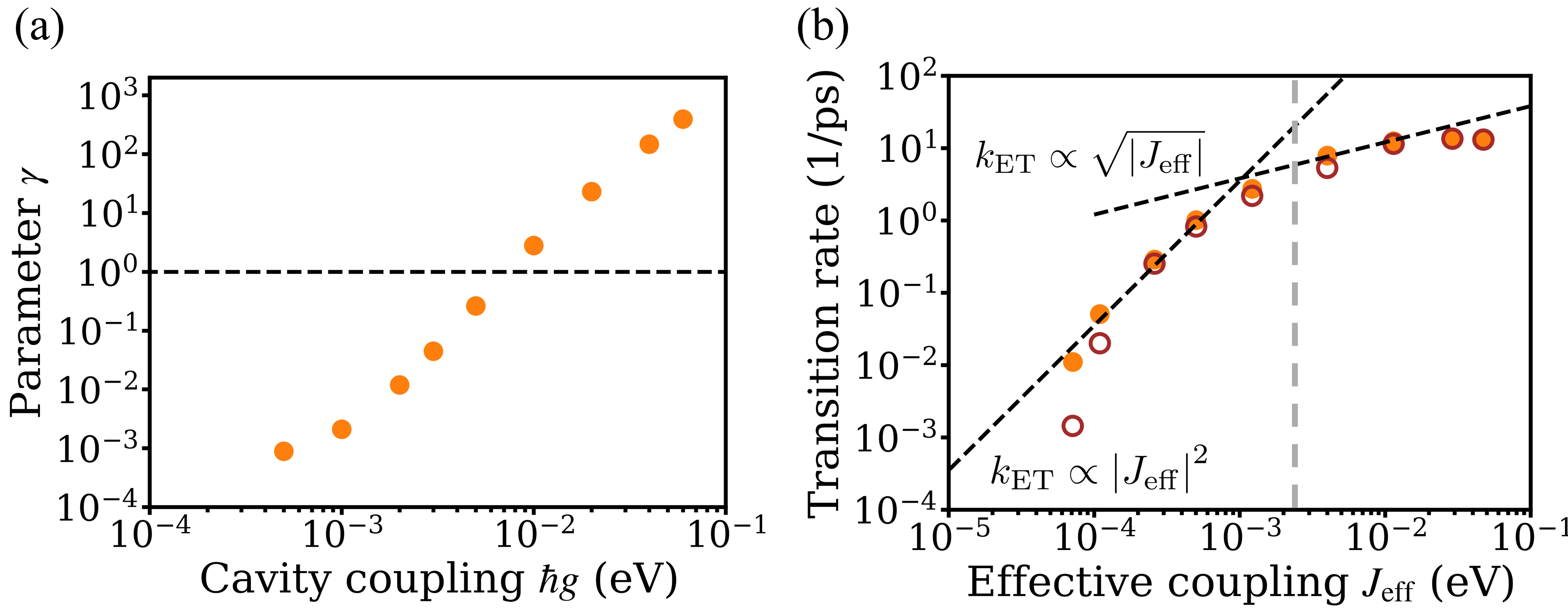}
\caption{\label{fig:fit_rate}(a) Plot of the nonadiabatic parameter $\gamma$ versus the QD-cavity coupling strength $\hbar g$ at 300K. The dashed line shows $\gamma=1$. (b) Energy transfer rate $k_{\text{ET}}$ as a function of effective coupling $J_{\text{eff}}$ for ideal cavity (orange dots) and lossy cavity (red hollow circles) at 300K. The black dashed lines show a quadratic and square root dependence and the gray dashed line corresponds to an effective coupling for $\gamma=1$.}
\end{figure*}

To better understand the transition from the overdamped to the coherent regime and make connections with previous works, we define a nonadiabatic parameter $\gamma$, given by:\citep{newton1984electron,balzani2001electron}
\begin{equation}
\gamma=\frac{\left|J_{{\rm eff}}\right|^{2}}{\hbar\omega_{{\rm nu}}}\sqrt{\frac{\pi^{3}}{\left(\lambda_{{\cal D}}+\lambda_{{\cal A}}\right)k_{B}T}},\label{eq:adiabatic parameter}
\end{equation}
where $\omega_{{\rm nu}}$
is the averaged effective phonon modes that contribute most to the system dynamics. The terms $\lambda_{{\cal D}}$ and $\lambda_{{\cal A}}$ represent
the average reorganization energies for the donor and acceptor, respectively,
defined as $\lambda_{{\cal D}/{\cal A}}=\sum_{n\in{\cal D}/{\cal A}}P_{n}\lambda_{n}$,
where $P_{n}$ is the Boltzmann population distribution of exciton
$|\psi_{n}\rangle$ and, as before, $\lambda_{n}=\frac{1}{2}\sum_{\alpha}\left(\tilde{V}_{nn}^{\alpha}\right)^{2}/\omega_{\alpha}^{2}$.
In Fig.~\ref{fig:fit_rate}(a) we plot the nonadiabatic parameter
$\gamma$ as a function of the QD-cavity coupling strength $\hbar g$.
For small coupling to the cavity mode, $\gamma\ll1$ and the system
is in the nonadiabatic regime, where the dynamics are characterized
by an exponential decay of populations. As the coupling to the cavity
mode increases, $\gamma$ also increases due to the increase in the
effective coupling between the two QDs. This parameter reaches and
surpasses $\gamma=1$ when $\hbar g=20$~meV, marking the transition
to an adiabatic regime. In this regime, the effective coupling strength
between the QDs becomes significantly larger than the couplings of
the donor and acceptor polaritonic states to phonons, leading to more
pronounced and sustained coherent dynamics.

In Fig.~\ref{fig:fit_rate}(b), we present the exciton transfer
rate as a function of the effective coupling for both ideal and lossy
cavities. For small value of $J_{\text{eff}}$, the transfer rate
scales with $J_{\text{eff}}^{2}$, consistent with Fermi's golden
rule. As $J_{\text{eff}}$ increases, the system transitions to the
adiabatic regime, and the transfer rate which is determined by decoherence times exhibits a more moderate dependence, fitting to a $\sqrt{J_{\text{eff}}}$ relationship. In this limit, the transfer rate was obtained by fitting the envelop function of $\rho_{{\cal D}}\left(t\right)$/$\rho_{{\cal A}}\left(t\right)$ to an exponential decay/rise. The gray dashed line in the figure marks the crossover coupling point, where the adiabatic parameter $\gamma=1$. In the lossy cavity, the energy transfer rate is inherently slower when compared to an ideal, lossless cavity. However, as the coupling to the cavity mode is increased, this discrepancy becomes less significant. This is because the rate of energy transfer accelerates, eventually becoming so rapid that it surpasses the cavity's inherent loss rate. As a result, the effects of the cavity's losses are effectively mitigated, rendering them negligible in the context of the swift energy transfer times that are achieved.

Before we conclude, we would like to mention that our calculations have been performed for a single QD pair and such large couplings required to drive the system into the adiabatic limit are typically achieved in plasmonic nanocavities, with complications arising from the coupling of excitons to near-fields.\cite{chikkaraddy2016single} While our approach ignores the role of near fields, it provides the guidelines that would be useful in future studies of QDs in nanocavities.

\section{Conclusion}
In this study, we investigated the dynamics of exciton energy transfer
between two nanocrystal quantum dots positioned within an optical
microcavity. We delved into this intricate phenomenon, examining how
the cavity photon mode influences the energy transfer process between
the quantum dots, essentially serving as a conduit. To achieve this,
we developed a model Hamiltonian, parameterized by atomistic calculations
of the excitonic level structure, phonon, exciton-phonon couplings,
and transition dipole moments of the semiconductor NCs within the
microcavity. To describe the dynamics comprehensively, we utilized
a quantum master equation and approximated its kernel to second order
in the exciton-phonon coupling (Redfield with memory), while also
performing a small polaron transformation to account for multiphonon
relaxation channels.

Our investigation revealed that by fine-tuning the coupling strength
between the cavity photon mode and the NCs, we can effectively modulate
the interaction between the donor and acceptor. Particularly noteworthy
is the identification of a critical transition from a nonadiabatic
to an adiabatic process, wherein the exciton energy transition rate
is accelerated and governed by decoherence times. This transition
holds significant implications for the controllability and efficiency
of energy transfer processes at the nanoscale. Overall, our findings
lay the groundwork for the design of innovative energy transfer systems
and nanophotonic devices, promising enhanced functionality and performance
in various applications.

\section*{Supplementary Material}
The supplementary material comprises nanocrystal configurations, 
simplification of polaron transformation, time--local non-Markovian Redfield equation, simplification of correlation function $\left\langle W_{nm}\left(t\right)W_{kl}\left(0\right)\right\rangle _{\text{B}}$, and derivation of the effective coupling for a 3-level system.

\begin{acknowledgments}
This work was supported by the U.S. Department of Energy, Office of
Science, Office of Basic Energy Sciences, Materials Sciences and Engineering
Division, under Contract No. DE461AC02-05-CH11231 within the Fundamentals
of Semiconductor Nanowire Program (KCPY23). Computational resources
were provided in part by the National Energy Research Scientific Computing
Center (NERSC), a U.S. Department of Energy Office of Science User
Facility operated under contract no. DEAC02-05CH11231. 
\end{acknowledgments}

\section*{Author Declarations}
\subsection*{Conflict of Interest Statement}
The authors have no conflicts to disclose.
\subsection*{Author Contributions}
Kaiyue Peng: Data curation (lead); Formal analysis (lead); Software (lead), Methodology (equal); Project administration (equal); Conceptualization (equal); Validation (lead); Writing – original draft (equal). Eran Rabani: Funding Acquisition (lead), Resources (lead), Supervision (lead), Writing/Review and Editing (lead), Methodology (equal), Project Administration (equal), Conceptualization (equal); Writing – original draft (equal).

\section*{Data Availability Statement}
The data that support the findings of this study are available
from the corresponding authors upon reasonable request.

\bibliography{reference_main}

\begin{thebibliography}{66}%
\makeatletter
\providecommand \@ifxundefined [1]{%
 \@ifx{#1\undefined}
}%
\providecommand \@ifnum [1]{%
 \ifnum #1\expandafter \@firstoftwo
 \else \expandafter \@secondoftwo
 \fi
}%
\providecommand \@ifx [1]{%
 \ifx #1\expandafter \@firstoftwo
 \else \expandafter \@secondoftwo
 \fi
}%
\providecommand \natexlab [1]{#1}%
\providecommand \enquote  [1]{``#1''}%
\providecommand \bibnamefont  [1]{#1}%
\providecommand \bibfnamefont [1]{#1}%
\providecommand \citenamefont [1]{#1}%
\providecommand \href@noop [0]{\@secondoftwo}%
\providecommand \href [0]{\begingroup \@sanitize@url \@href}%
\providecommand \@href[1]{\@@startlink{#1}\@@href}%
\providecommand \@@href[1]{\endgroup#1\@@endlink}%
\providecommand \@sanitize@url [0]{\catcode `\\12\catcode `\$12\catcode `\&12\catcode `\#12\catcode `\^12\catcode `\_12\catcode `\%12\relax}%
\providecommand \@@startlink[1]{}%
\providecommand \@@endlink[0]{}%
\providecommand \url  [0]{\begingroup\@sanitize@url \@url }%
\providecommand \@url [1]{\endgroup\@href {#1}{\urlprefix }}%
\providecommand \urlprefix  [0]{URL }%
\providecommand \Eprint [0]{\href }%
\providecommand \doibase [0]{https://doi.org/}%
\providecommand \selectlanguage [0]{\@gobble}%
\providecommand \bibinfo  [0]{\@secondoftwo}%
\providecommand \bibfield  [0]{\@secondoftwo}%
\providecommand \translation [1]{[#1]}%
\providecommand \BibitemOpen [0]{}%
\providecommand \bibitemStop [0]{}%
\providecommand \bibitemNoStop [0]{.\EOS\space}%
\providecommand \EOS [0]{\spacefactor3000\relax}%
\providecommand \BibitemShut  [1]{\csname bibitem#1\endcsname}%
\let\auto@bib@innerbib\@empty
\bibitem [{\citenamefont {Choi}\ \emph {et~al.}(2012)\citenamefont {Choi}, \citenamefont {Jin}, \citenamefont {Bang},\ and\ \citenamefont {Kim}}]{choi2012layer}%
  \BibitemOpen
  \bibfield  {author} {\bibinfo {author} {\bibfnamefont {S.}~\bibnamefont {Choi}}, \bibinfo {author} {\bibfnamefont {H.}~\bibnamefont {Jin}}, \bibinfo {author} {\bibfnamefont {J.}~\bibnamefont {Bang}},\ and\ \bibinfo {author} {\bibfnamefont {S.}~\bibnamefont {Kim}},\ }\bibfield  {title} {\enquote {\bibinfo {title} {Layer-by-layer quantum dot assemblies for the enhanced energy transfers and their applications toward efficient solar cells},}\ }\href@noop {} {\bibfield  {journal} {\bibinfo  {journal} {J. Phys. Chem. Lett.}\ }\textbf {\bibinfo {volume} {3}},\ \bibinfo {pages} {3442--3447} (\bibinfo {year} {2012})}\BibitemShut {NoStop}%
\bibitem [{\citenamefont {Almutlaq}\ \emph {et~al.}(2024)\citenamefont {Almutlaq}, \citenamefont {Liu}, \citenamefont {Mir}, \citenamefont {Sabatini}, \citenamefont {Englund}, \citenamefont {Bakr},\ and\ \citenamefont {Sargent}}]{almutlaq2024engineering}%
  \BibitemOpen
  \bibfield  {author} {\bibinfo {author} {\bibfnamefont {J.}~\bibnamefont {Almutlaq}}, \bibinfo {author} {\bibfnamefont {Y.}~\bibnamefont {Liu}}, \bibinfo {author} {\bibfnamefont {W.~J.}\ \bibnamefont {Mir}}, \bibinfo {author} {\bibfnamefont {R.~P.}\ \bibnamefont {Sabatini}}, \bibinfo {author} {\bibfnamefont {D.}~\bibnamefont {Englund}}, \bibinfo {author} {\bibfnamefont {O.~M.}\ \bibnamefont {Bakr}},\ and\ \bibinfo {author} {\bibfnamefont {E.~H.}\ \bibnamefont {Sargent}},\ }\bibfield  {title} {\enquote {\bibinfo {title} {Engineering colloidal semiconductor nanocrystals for quantum information processing},}\ }\href@noop {} {\bibfield  {journal} {\bibinfo  {journal} {Nat. Nanotechnol.}\ ,\ \bibinfo {pages} {1--10}} (\bibinfo {year} {2024})}\BibitemShut {NoStop}%
\bibitem [{\citenamefont {Alivisatos}(1996)}]{alivisatos1996perspectives}%
  \BibitemOpen
  \bibfield  {author} {\bibinfo {author} {\bibfnamefont {A.~P.}\ \bibnamefont {Alivisatos}},\ }\bibfield  {title} {\enquote {\bibinfo {title} {Perspectives on the physical chemistry of semiconductor nanocrystals},}\ }\href@noop {} {\bibfield  {journal} {\bibinfo  {journal} {J. Phys. Chem.}\ }\textbf {\bibinfo {volume} {100}},\ \bibinfo {pages} {13226--13239} (\bibinfo {year} {1996})}\BibitemShut {NoStop}%
\bibitem [{\citenamefont {Efros}\ and\ \citenamefont {Rosen}(2000)}]{efros2000electronic}%
  \BibitemOpen
  \bibfield  {author} {\bibinfo {author} {\bibfnamefont {A.~L.}\ \bibnamefont {Efros}}\ and\ \bibinfo {author} {\bibfnamefont {M.}~\bibnamefont {Rosen}},\ }\bibfield  {title} {\enquote {\bibinfo {title} {The electronic structure of semiconductor nanocrystals},}\ }\href@noop {} {\bibfield  {journal} {\bibinfo  {journal} {Annu. Rev. Mater. Sci.}\ }\textbf {\bibinfo {volume} {30}},\ \bibinfo {pages} {475--521} (\bibinfo {year} {2000})}\BibitemShut {NoStop}%
\bibitem [{\citenamefont {Gomez}, \citenamefont {Califano},\ and\ \citenamefont {Mulvaney}(2006)}]{gomez2006optical}%
  \BibitemOpen
  \bibfield  {author} {\bibinfo {author} {\bibfnamefont {D.~E.}\ \bibnamefont {Gomez}}, \bibinfo {author} {\bibfnamefont {M.}~\bibnamefont {Califano}},\ and\ \bibinfo {author} {\bibfnamefont {P.}~\bibnamefont {Mulvaney}},\ }\bibfield  {title} {\enquote {\bibinfo {title} {Optical properties of single semiconductor nanocrystals},}\ }\href@noop {} {\bibfield  {journal} {\bibinfo  {journal} {Phys. Chem. Chem. Phys.}\ }\textbf {\bibinfo {volume} {8}},\ \bibinfo {pages} {4989--5011} (\bibinfo {year} {2006})}\BibitemShut {NoStop}%
\bibitem [{\citenamefont {Klimov}(2007)}]{klimov2007spectral}%
  \BibitemOpen
  \bibfield  {author} {\bibinfo {author} {\bibfnamefont {V.~I.}\ \bibnamefont {Klimov}},\ }\bibfield  {title} {\enquote {\bibinfo {title} {Spectral and dynamical properties of multiexcitons in semiconductor nanocrystals},}\ }\href@noop {} {\bibfield  {journal} {\bibinfo  {journal} {Annu. Rev. Phys. Chem.}\ }\textbf {\bibinfo {volume} {58}},\ \bibinfo {pages} {635--673} (\bibinfo {year} {2007})}\BibitemShut {NoStop}%
\bibitem [{\citenamefont {Kodaimati}\ \emph {et~al.}(2018)\citenamefont {Kodaimati}, \citenamefont {Lian}, \citenamefont {Schatz},\ and\ \citenamefont {Weiss}}]{kodaimati2018energy}%
  \BibitemOpen
  \bibfield  {author} {\bibinfo {author} {\bibfnamefont {M.~S.}\ \bibnamefont {Kodaimati}}, \bibinfo {author} {\bibfnamefont {S.}~\bibnamefont {Lian}}, \bibinfo {author} {\bibfnamefont {G.~C.}\ \bibnamefont {Schatz}},\ and\ \bibinfo {author} {\bibfnamefont {E.~A.}\ \bibnamefont {Weiss}},\ }\bibfield  {title} {\enquote {\bibinfo {title} {Energy transfer-enhanced photocatalytic reduction of protons within quantum dot light-harvesting--catalyst assemblies},}\ }\href@noop {} {\bibfield  {journal} {\bibinfo  {journal} {Proc. Natl. Acad. Sci. U. S. A.}\ }\textbf {\bibinfo {volume} {115}},\ \bibinfo {pages} {8290--8295} (\bibinfo {year} {2018})}\BibitemShut {NoStop}%
\bibitem [{\citenamefont {Roy}\ \emph {et~al.}(2020)\citenamefont {Roy}, \citenamefont {Devatha}, \citenamefont {Roy}, \citenamefont {Rao},\ and\ \citenamefont {Pillai}}]{roy2020electrostatically}%
  \BibitemOpen
  \bibfield  {author} {\bibinfo {author} {\bibfnamefont {P.}~\bibnamefont {Roy}}, \bibinfo {author} {\bibfnamefont {G.}~\bibnamefont {Devatha}}, \bibinfo {author} {\bibfnamefont {S.}~\bibnamefont {Roy}}, \bibinfo {author} {\bibfnamefont {A.}~\bibnamefont {Rao}},\ and\ \bibinfo {author} {\bibfnamefont {P.~P.}\ \bibnamefont {Pillai}},\ }\bibfield  {title} {\enquote {\bibinfo {title} {Electrostatically driven resonance energy transfer in an all-quantum dot based donor--acceptor system},}\ }\href@noop {} {\bibfield  {journal} {\bibinfo  {journal} {J. Phys. Chem. Lett.}\ }\textbf {\bibinfo {volume} {11}},\ \bibinfo {pages} {5354--5360} (\bibinfo {year} {2020})}\BibitemShut {NoStop}%
\bibitem [{\citenamefont {Kodaimati}\ \emph {et~al.}(2017)\citenamefont {Kodaimati}, \citenamefont {Wang}, \citenamefont {Chapman}, \citenamefont {Schatz},\ and\ \citenamefont {Weiss}}]{kodaimati2017distance}%
  \BibitemOpen
  \bibfield  {author} {\bibinfo {author} {\bibfnamefont {M.~S.}\ \bibnamefont {Kodaimati}}, \bibinfo {author} {\bibfnamefont {C.}~\bibnamefont {Wang}}, \bibinfo {author} {\bibfnamefont {C.}~\bibnamefont {Chapman}}, \bibinfo {author} {\bibfnamefont {G.~C.}\ \bibnamefont {Schatz}},\ and\ \bibinfo {author} {\bibfnamefont {E.~A.}\ \bibnamefont {Weiss}},\ }\bibfield  {title} {\enquote {\bibinfo {title} {Distance-dependence of interparticle energy transfer in the near-infrared within electrostatic assemblies of pbs quantum dots},}\ }\href@noop {} {\bibfield  {journal} {\bibinfo  {journal} {ACS nano}\ }\textbf {\bibinfo {volume} {11}},\ \bibinfo {pages} {5041--5050} (\bibinfo {year} {2017})}\BibitemShut {NoStop}%
\bibitem [{\citenamefont {Scholes}, \citenamefont {Jordanides},\ and\ \citenamefont {Fleming}(2001)}]{scholes2001adapting}%
  \BibitemOpen
  \bibfield  {author} {\bibinfo {author} {\bibfnamefont {G.~D.}\ \bibnamefont {Scholes}}, \bibinfo {author} {\bibfnamefont {X.~J.}\ \bibnamefont {Jordanides}},\ and\ \bibinfo {author} {\bibfnamefont {G.~R.}\ \bibnamefont {Fleming}},\ }\bibfield  {title} {\enquote {\bibinfo {title} {Adapting the f{\"o}rster theory of energy transfer for modeling dynamics in aggregated molecular assemblies},}\ }\href@noop {} {\bibfield  {journal} {\bibinfo  {journal} {J. Phys. Chem. B}\ }\textbf {\bibinfo {volume} {105}},\ \bibinfo {pages} {1640--1651} (\bibinfo {year} {2001})}\BibitemShut {NoStop}%
\bibitem [{\citenamefont {Barth}\ \emph {et~al.}(2022)\citenamefont {Barth}, \citenamefont {Opanasyuk}, \citenamefont {Peulen}, \citenamefont {Felekyan}, \citenamefont {Kalinin}, \citenamefont {Sanabria},\ and\ \citenamefont {Seidel}}]{barth2022unraveling}%
  \BibitemOpen
  \bibfield  {author} {\bibinfo {author} {\bibfnamefont {A.}~\bibnamefont {Barth}}, \bibinfo {author} {\bibfnamefont {O.}~\bibnamefont {Opanasyuk}}, \bibinfo {author} {\bibfnamefont {T.-O.}\ \bibnamefont {Peulen}}, \bibinfo {author} {\bibfnamefont {S.}~\bibnamefont {Felekyan}}, \bibinfo {author} {\bibfnamefont {S.}~\bibnamefont {Kalinin}}, \bibinfo {author} {\bibfnamefont {H.}~\bibnamefont {Sanabria}},\ and\ \bibinfo {author} {\bibfnamefont {C.~A.~M.}\ \bibnamefont {Seidel}},\ }\bibfield  {title} {\enquote {\bibinfo {title} {{Unraveling multi-state molecular dynamics in single-molecule FRET experiments. I. Theory of FRET-lines}},}\ }\href@noop {} {\bibfield  {journal} {\bibinfo  {journal} {J. Chem. Phys.}\ }\textbf {\bibinfo {volume} {156}},\ \bibinfo {pages} {141501} (\bibinfo {year} {2022})}\BibitemShut {NoStop}%
\bibitem [{\citenamefont {Guo}\ \emph {et~al.}(2018)\citenamefont {Guo}, \citenamefont {Qiu}, \citenamefont {Mingoes}, \citenamefont {Deschamps}, \citenamefont {Susumu}, \citenamefont {Medintz},\ and\ \citenamefont {Hildebrandt}}]{guo2018conformational}%
  \BibitemOpen
  \bibfield  {author} {\bibinfo {author} {\bibfnamefont {J.}~\bibnamefont {Guo}}, \bibinfo {author} {\bibfnamefont {X.}~\bibnamefont {Qiu}}, \bibinfo {author} {\bibfnamefont {C.}~\bibnamefont {Mingoes}}, \bibinfo {author} {\bibfnamefont {J.~R.}\ \bibnamefont {Deschamps}}, \bibinfo {author} {\bibfnamefont {K.}~\bibnamefont {Susumu}}, \bibinfo {author} {\bibfnamefont {I.~L.}\ \bibnamefont {Medintz}},\ and\ \bibinfo {author} {\bibfnamefont {N.}~\bibnamefont {Hildebrandt}},\ }\bibfield  {title} {\enquote {\bibinfo {title} {Conformational details of quantum dot-dna resolved by forster resonance energy transfer lifetime nanoruler},}\ }\href@noop {} {\bibfield  {journal} {\bibinfo  {journal} {ACS nano}\ }\textbf {\bibinfo {volume} {13}},\ \bibinfo {pages} {505--514} (\bibinfo {year} {2018})}\BibitemShut {NoStop}%
\bibitem [{\citenamefont {Jares-Erijman}\ and\ \citenamefont {Jovin}(2003)}]{jares2003fret}%
  \BibitemOpen
  \bibfield  {author} {\bibinfo {author} {\bibfnamefont {E.~A.}\ \bibnamefont {Jares-Erijman}}\ and\ \bibinfo {author} {\bibfnamefont {T.~M.}\ \bibnamefont {Jovin}},\ }\bibfield  {title} {\enquote {\bibinfo {title} {Fret imaging},}\ }\href@noop {} {\bibfield  {journal} {\bibinfo  {journal} {Nat. Biotechnol.}\ }\textbf {\bibinfo {volume} {21}},\ \bibinfo {pages} {1387--1395} (\bibinfo {year} {2003})}\BibitemShut {NoStop}%
\bibitem [{\citenamefont {Kim}, \citenamefont {Kim},\ and\ \citenamefont {Lee}(2015)}]{kim2015real}%
  \BibitemOpen
  \bibfield  {author} {\bibinfo {author} {\bibfnamefont {J.-Y.}\ \bibnamefont {Kim}}, \bibinfo {author} {\bibfnamefont {C.}~\bibnamefont {Kim}},\ and\ \bibinfo {author} {\bibfnamefont {N.~K.}\ \bibnamefont {Lee}},\ }\bibfield  {title} {\enquote {\bibinfo {title} {Real-time submillisecond single-molecule fret dynamics of freely diffusing molecules with liposome tethering},}\ }\href@noop {} {\bibfield  {journal} {\bibinfo  {journal} {Nat. Commun.}\ }\textbf {\bibinfo {volume} {6}},\ \bibinfo {pages} {6992} (\bibinfo {year} {2015})}\BibitemShut {NoStop}%
\bibitem [{\citenamefont {Sun}\ \emph {et~al.}(2011)\citenamefont {Sun}, \citenamefont {Wallrabe}, \citenamefont {Seo},\ and\ \citenamefont {Periasamy}}]{sun2011fret}%
  \BibitemOpen
  \bibfield  {author} {\bibinfo {author} {\bibfnamefont {Y.}~\bibnamefont {Sun}}, \bibinfo {author} {\bibfnamefont {H.}~\bibnamefont {Wallrabe}}, \bibinfo {author} {\bibfnamefont {S.-A.}\ \bibnamefont {Seo}},\ and\ \bibinfo {author} {\bibfnamefont {A.}~\bibnamefont {Periasamy}},\ }\bibfield  {title} {\enquote {\bibinfo {title} {Fret microscopy in 2010: the legacy of theodor f{\"o}rster on the 100th anniversary of his birth},}\ }\href@noop {} {\bibfield  {journal} {\bibinfo  {journal} {ChemPhysChem}\ }\textbf {\bibinfo {volume} {12}},\ \bibinfo {pages} {462--474} (\bibinfo {year} {2011})}\BibitemShut {NoStop}%
\bibitem [{\citenamefont {Zhong}\ \emph {et~al.}(2023)\citenamefont {Zhong}, \citenamefont {Nguyen}, \citenamefont {Do}, \citenamefont {Tan}, \citenamefont {Knoester},\ and\ \citenamefont {Jansen}}]{zhong2023efficient}%
  \BibitemOpen
  \bibfield  {author} {\bibinfo {author} {\bibfnamefont {K.}~\bibnamefont {Zhong}}, \bibinfo {author} {\bibfnamefont {H.~L.}\ \bibnamefont {Nguyen}}, \bibinfo {author} {\bibfnamefont {T.~N.}\ \bibnamefont {Do}}, \bibinfo {author} {\bibfnamefont {H.-S.}\ \bibnamefont {Tan}}, \bibinfo {author} {\bibfnamefont {J.}~\bibnamefont {Knoester}},\ and\ \bibinfo {author} {\bibfnamefont {T.~L.~C.}\ \bibnamefont {Jansen}},\ }\bibfield  {title} {\enquote {\bibinfo {title} {{An efficient time-domain implementation of the multichromophoric Förster resonant energy transfer method}},}\ }\href@noop {} {\bibfield  {journal} {\bibinfo  {journal} {J. Chem. Phys.}\ }\textbf {\bibinfo {volume} {158}},\ \bibinfo {pages} {064103} (\bibinfo {year} {2023})}\BibitemShut {NoStop}%
\bibitem [{\citenamefont {Baer}\ and\ \citenamefont {Rabani}(2008)}]{baer2008theory}%
  \BibitemOpen
  \bibfield  {author} {\bibinfo {author} {\bibfnamefont {R.}~\bibnamefont {Baer}}\ and\ \bibinfo {author} {\bibfnamefont {E.}~\bibnamefont {Rabani}},\ }\bibfield  {title} {\enquote {\bibinfo {title} {{Theory of resonance energy transfer involving nanocrystals: The role of high multipoles}},}\ }\href@noop {} {\bibfield  {journal} {\bibinfo  {journal} {J. Chem. Phys.}\ }\textbf {\bibinfo {volume} {128}},\ \bibinfo {pages} {184710} (\bibinfo {year} {2008})}\BibitemShut {NoStop}%
\bibitem [{\citenamefont {Loiudice}, \citenamefont {Saris},\ and\ \citenamefont {Buonsanti}(2020)}]{loiudice2020tunable}%
  \BibitemOpen
  \bibfield  {author} {\bibinfo {author} {\bibfnamefont {A.}~\bibnamefont {Loiudice}}, \bibinfo {author} {\bibfnamefont {S.}~\bibnamefont {Saris}},\ and\ \bibinfo {author} {\bibfnamefont {R.}~\bibnamefont {Buonsanti}},\ }\bibfield  {title} {\enquote {\bibinfo {title} {Tunable metal oxide shell as a spacer to study energy transfer in semiconductor nanocrystals},}\ }\href@noop {} {\bibfield  {journal} {\bibinfo  {journal} {J. Phys. Chem. Lett.}\ }\textbf {\bibinfo {volume} {11}},\ \bibinfo {pages} {3430--3435} (\bibinfo {year} {2020})}\BibitemShut {NoStop}%
\bibitem [{\citenamefont {Govorov}, \citenamefont {Lee},\ and\ \citenamefont {Kotov}(2007)}]{govorov2007theory}%
  \BibitemOpen
  \bibfield  {author} {\bibinfo {author} {\bibfnamefont {A.~O.}\ \bibnamefont {Govorov}}, \bibinfo {author} {\bibfnamefont {J.}~\bibnamefont {Lee}},\ and\ \bibinfo {author} {\bibfnamefont {N.~A.}\ \bibnamefont {Kotov}},\ }\bibfield  {title} {\enquote {\bibinfo {title} {Theory of plasmon-enhanced f{\"o}rster energy transfer in optically excited semiconductor and metal nanoparticles},}\ }\href@noop {} {\bibfield  {journal} {\bibinfo  {journal} {Phys. Rev. B}\ }\textbf {\bibinfo {volume} {76}},\ \bibinfo {pages} {125308} (\bibinfo {year} {2007})}\BibitemShut {NoStop}%
\bibitem [{\citenamefont {Schachenmayer}\ \emph {et~al.}(2015)\citenamefont {Schachenmayer}, \citenamefont {Genes}, \citenamefont {Tignone},\ and\ \citenamefont {Pupillo}}]{schachenmayer2015cavity}%
  \BibitemOpen
  \bibfield  {author} {\bibinfo {author} {\bibfnamefont {J.}~\bibnamefont {Schachenmayer}}, \bibinfo {author} {\bibfnamefont {C.}~\bibnamefont {Genes}}, \bibinfo {author} {\bibfnamefont {E.}~\bibnamefont {Tignone}},\ and\ \bibinfo {author} {\bibfnamefont {G.}~\bibnamefont {Pupillo}},\ }\bibfield  {title} {\enquote {\bibinfo {title} {Cavity-enhanced transport of excitons},}\ }\href@noop {} {\bibfield  {journal} {\bibinfo  {journal} {Phys. Rev. Lett.}\ }\textbf {\bibinfo {volume} {114}},\ \bibinfo {pages} {196403} (\bibinfo {year} {2015})}\BibitemShut {NoStop}%
\bibitem [{\citenamefont {Wang}, \citenamefont {Hertzog},\ and\ \citenamefont {B{\"o}rjesson}(2021)}]{wang2021polariton}%
  \BibitemOpen
  \bibfield  {author} {\bibinfo {author} {\bibfnamefont {M.}~\bibnamefont {Wang}}, \bibinfo {author} {\bibfnamefont {M.}~\bibnamefont {Hertzog}},\ and\ \bibinfo {author} {\bibfnamefont {K.}~\bibnamefont {B{\"o}rjesson}},\ }\bibfield  {title} {\enquote {\bibinfo {title} {Polariton-assisted excitation energy channeling in organic heterojunctions},}\ }\href@noop {} {\bibfield  {journal} {\bibinfo  {journal} {Nat. Commun.}\ }\textbf {\bibinfo {volume} {12}},\ \bibinfo {pages} {1874} (\bibinfo {year} {2021})}\BibitemShut {NoStop}%
\bibitem [{\citenamefont {Chowdhury}, \citenamefont {Zhang},\ and\ \citenamefont {Beratan}(2022)}]{chowdhury2022interference}%
  \BibitemOpen
  \bibfield  {author} {\bibinfo {author} {\bibfnamefont {S.~N.}\ \bibnamefont {Chowdhury}}, \bibinfo {author} {\bibfnamefont {P.}~\bibnamefont {Zhang}},\ and\ \bibinfo {author} {\bibfnamefont {D.~N.}\ \bibnamefont {Beratan}},\ }\bibfield  {title} {\enquote {\bibinfo {title} {Interference between molecular and photon field-mediated electron transfer coupling pathways in cavities},}\ }\href@noop {} {\bibfield  {journal} {\bibinfo  {journal} {J. Phys. Chem. Lett.}\ }\textbf {\bibinfo {volume} {13}},\ \bibinfo {pages} {9822--9828} (\bibinfo {year} {2022})}\BibitemShut {NoStop}%
\bibitem [{\citenamefont {Albinsson}\ and\ \citenamefont {M{\aa}rtensson}(2010)}]{albinsson2010excitation}%
  \BibitemOpen
  \bibfield  {author} {\bibinfo {author} {\bibfnamefont {B.}~\bibnamefont {Albinsson}}\ and\ \bibinfo {author} {\bibfnamefont {J.}~\bibnamefont {M{\aa}rtensson}},\ }\bibfield  {title} {\enquote {\bibinfo {title} {Excitation energy transfer in donor--bridge--acceptor systems},}\ }\href@noop {} {\bibfield  {journal} {\bibinfo  {journal} {Phys. Chem. Chem. Phys.}\ }\textbf {\bibinfo {volume} {12}},\ \bibinfo {pages} {7338--7351} (\bibinfo {year} {2010})}\BibitemShut {NoStop}%
\bibitem [{\citenamefont {Xu}\ \emph {et~al.}(2023)\citenamefont {Xu}, \citenamefont {Mandal}, \citenamefont {Baxter}, \citenamefont {Cheng}, \citenamefont {Lee}, \citenamefont {Su}, \citenamefont {Liu}, \citenamefont {Reichman},\ and\ \citenamefont {Delor}}]{xu2023ultrafast}%
  \BibitemOpen
  \bibfield  {author} {\bibinfo {author} {\bibfnamefont {D.}~\bibnamefont {Xu}}, \bibinfo {author} {\bibfnamefont {A.}~\bibnamefont {Mandal}}, \bibinfo {author} {\bibfnamefont {J.~M.}\ \bibnamefont {Baxter}}, \bibinfo {author} {\bibfnamefont {S.-W.}\ \bibnamefont {Cheng}}, \bibinfo {author} {\bibfnamefont {I.}~\bibnamefont {Lee}}, \bibinfo {author} {\bibfnamefont {H.}~\bibnamefont {Su}}, \bibinfo {author} {\bibfnamefont {S.}~\bibnamefont {Liu}}, \bibinfo {author} {\bibfnamefont {D.~R.}\ \bibnamefont {Reichman}},\ and\ \bibinfo {author} {\bibfnamefont {M.}~\bibnamefont {Delor}},\ }\bibfield  {title} {\enquote {\bibinfo {title} {Ultrafast imaging of polariton propagation and interactions},}\ }\href@noop {} {\bibfield  {journal} {\bibinfo  {journal} {Nat. Commun.}\ }\textbf {\bibinfo {volume} {14}},\ \bibinfo {pages} {3881} (\bibinfo {year} {2023})}\BibitemShut {NoStop}%
\bibitem [{\citenamefont {Feist}\ and\ \citenamefont {Garcia-Vidal}(2015)}]{feist2015extraordinary}%
  \BibitemOpen
  \bibfield  {author} {\bibinfo {author} {\bibfnamefont {J.}~\bibnamefont {Feist}}\ and\ \bibinfo {author} {\bibfnamefont {F.~J.}\ \bibnamefont {Garcia-Vidal}},\ }\bibfield  {title} {\enquote {\bibinfo {title} {Extraordinary exciton conductance induced by strong coupling},}\ }\href@noop {} {\bibfield  {journal} {\bibinfo  {journal} {Phys. Rev. Lett.}\ }\textbf {\bibinfo {volume} {114}},\ \bibinfo {pages} {196402} (\bibinfo {year} {2015})}\BibitemShut {NoStop}%
\bibitem [{\citenamefont {Zhong}\ \emph {et~al.}(2017)\citenamefont {Zhong}, \citenamefont {Chervy}, \citenamefont {Zhang}, \citenamefont {Thomas}, \citenamefont {George}, \citenamefont {Genet}, \citenamefont {Hutchison},\ and\ \citenamefont {Ebbesen}}]{zhong2017energy}%
  \BibitemOpen
  \bibfield  {author} {\bibinfo {author} {\bibfnamefont {X.}~\bibnamefont {Zhong}}, \bibinfo {author} {\bibfnamefont {T.}~\bibnamefont {Chervy}}, \bibinfo {author} {\bibfnamefont {L.}~\bibnamefont {Zhang}}, \bibinfo {author} {\bibfnamefont {A.}~\bibnamefont {Thomas}}, \bibinfo {author} {\bibfnamefont {J.}~\bibnamefont {George}}, \bibinfo {author} {\bibfnamefont {C.}~\bibnamefont {Genet}}, \bibinfo {author} {\bibfnamefont {J.~A.}\ \bibnamefont {Hutchison}},\ and\ \bibinfo {author} {\bibfnamefont {T.~W.}\ \bibnamefont {Ebbesen}},\ }\bibfield  {title} {\enquote {\bibinfo {title} {Energy transfer between spatially separated entangled molecules},}\ }\href@noop {} {\bibfield  {journal} {\bibinfo  {journal} {Angew. Chem.}\ }\textbf {\bibinfo {volume} {129}},\ \bibinfo {pages} {9162--9166} (\bibinfo {year} {2017})}\BibitemShut {NoStop}%
\bibitem [{\citenamefont {Weight}, \citenamefont {Krauss},\ and\ \citenamefont {Huo}(2023)}]{weight2023investigating}%
  \BibitemOpen
  \bibfield  {author} {\bibinfo {author} {\bibfnamefont {B.~M.}\ \bibnamefont {Weight}}, \bibinfo {author} {\bibfnamefont {T.~D.}\ \bibnamefont {Krauss}},\ and\ \bibinfo {author} {\bibfnamefont {P.}~\bibnamefont {Huo}},\ }\bibfield  {title} {\enquote {\bibinfo {title} {Investigating molecular exciton polaritons using ab initio cavity quantum electrodynamics},}\ }\href@noop {} {\bibfield  {journal} {\bibinfo  {journal} {J. Phys. Chem. Lett.}\ }\textbf {\bibinfo {volume} {14}},\ \bibinfo {pages} {5901--5913} (\bibinfo {year} {2023})}\BibitemShut {NoStop}%
\bibitem [{\citenamefont {Du}\ \emph {et~al.}(2018)\citenamefont {Du}, \citenamefont {Mart{\'\i}nez-Mart{\'\i}nez}, \citenamefont {Ribeiro}, \citenamefont {Hu}, \citenamefont {Menon},\ and\ \citenamefont {Yuen-Zhou}}]{du2018theory}%
  \BibitemOpen
  \bibfield  {author} {\bibinfo {author} {\bibfnamefont {M.}~\bibnamefont {Du}}, \bibinfo {author} {\bibfnamefont {L.~A.}\ \bibnamefont {Mart{\'\i}nez-Mart{\'\i}nez}}, \bibinfo {author} {\bibfnamefont {R.~F.}\ \bibnamefont {Ribeiro}}, \bibinfo {author} {\bibfnamefont {Z.}~\bibnamefont {Hu}}, \bibinfo {author} {\bibfnamefont {V.~M.}\ \bibnamefont {Menon}},\ and\ \bibinfo {author} {\bibfnamefont {J.}~\bibnamefont {Yuen-Zhou}},\ }\bibfield  {title} {\enquote {\bibinfo {title} {Theory for polariton-assisted remote energy transfer},}\ }\href@noop {} {\bibfield  {journal} {\bibinfo  {journal} {Chem. Sci.}\ }\textbf {\bibinfo {volume} {9}},\ \bibinfo {pages} {6659--6669} (\bibinfo {year} {2018})}\BibitemShut {NoStop}%
\bibitem [{\citenamefont {Kimura}(2009)}]{kimura2009general}%
  \BibitemOpen
  \bibfield  {author} {\bibinfo {author} {\bibfnamefont {A.}~\bibnamefont {Kimura}},\ }\bibfield  {title} {\enquote {\bibinfo {title} {General theory of excitation energy transfer in donor-mediator-acceptor systems},}\ }\href@noop {} {\bibfield  {journal} {\bibinfo  {journal} {J. Chem. Phys.}\ }\textbf {\bibinfo {volume} {130}},\ \bibinfo {pages} {154103} (\bibinfo {year} {2009})}\BibitemShut {NoStop}%
\bibitem [{\citenamefont {S{\'a}ez-Bl{\'a}zquez}\ \emph {et~al.}(2018)\citenamefont {S{\'a}ez-Bl{\'a}zquez}, \citenamefont {Feist}, \citenamefont {Fern{\'a}ndez-Dom{\'\i}nguez},\ and\ \citenamefont {Garc{\'\i}a-Vidal}}]{saez2018organic}%
  \BibitemOpen
  \bibfield  {author} {\bibinfo {author} {\bibfnamefont {R.}~\bibnamefont {S{\'a}ez-Bl{\'a}zquez}}, \bibinfo {author} {\bibfnamefont {J.}~\bibnamefont {Feist}}, \bibinfo {author} {\bibfnamefont {A.~I.}\ \bibnamefont {Fern{\'a}ndez-Dom{\'\i}nguez}},\ and\ \bibinfo {author} {\bibfnamefont {F.}~\bibnamefont {Garc{\'\i}a-Vidal}},\ }\bibfield  {title} {\enquote {\bibinfo {title} {Organic polaritons enable local vibrations to drive long-range energy transfer},}\ }\href@noop {} {\bibfield  {journal} {\bibinfo  {journal} {Phys. Rev. B}\ }\textbf {\bibinfo {volume} {97}},\ \bibinfo {pages} {241407} (\bibinfo {year} {2018})}\BibitemShut {NoStop}%
\bibitem [{\citenamefont {Reitz}, \citenamefont {Mineo},\ and\ \citenamefont {Genes}(2018)}]{reitz2018energy}%
  \BibitemOpen
  \bibfield  {author} {\bibinfo {author} {\bibfnamefont {M.}~\bibnamefont {Reitz}}, \bibinfo {author} {\bibfnamefont {F.}~\bibnamefont {Mineo}},\ and\ \bibinfo {author} {\bibfnamefont {C.}~\bibnamefont {Genes}},\ }\bibfield  {title} {\enquote {\bibinfo {title} {Energy transfer and correlations in cavity-embedded donor-acceptor configurations},}\ }\href@noop {} {\bibfield  {journal} {\bibinfo  {journal} {Sci. Rep.}\ }\textbf {\bibinfo {volume} {8}},\ \bibinfo {pages} {9050} (\bibinfo {year} {2018})}\BibitemShut {NoStop}%
\bibitem [{\citenamefont {Coles}\ \emph {et~al.}(2014)\citenamefont {Coles}, \citenamefont {Somaschi}, \citenamefont {Michetti}, \citenamefont {Clark}, \citenamefont {Lagoudakis}, \citenamefont {Savvidis},\ and\ \citenamefont {Lidzey}}]{coles2014polariton}%
  \BibitemOpen
  \bibfield  {author} {\bibinfo {author} {\bibfnamefont {D.~M.}\ \bibnamefont {Coles}}, \bibinfo {author} {\bibfnamefont {N.}~\bibnamefont {Somaschi}}, \bibinfo {author} {\bibfnamefont {P.}~\bibnamefont {Michetti}}, \bibinfo {author} {\bibfnamefont {C.}~\bibnamefont {Clark}}, \bibinfo {author} {\bibfnamefont {P.~G.}\ \bibnamefont {Lagoudakis}}, \bibinfo {author} {\bibfnamefont {P.~G.}\ \bibnamefont {Savvidis}},\ and\ \bibinfo {author} {\bibfnamefont {D.~G.}\ \bibnamefont {Lidzey}},\ }\bibfield  {title} {\enquote {\bibinfo {title} {Polariton-mediated energy transfer between organic dyes in a strongly coupled optical microcavity},}\ }\href@noop {} {\bibfield  {journal} {\bibinfo  {journal} {Nat. Mater.}\ }\textbf {\bibinfo {volume} {13}},\ \bibinfo {pages} {712--719} (\bibinfo {year} {2014})}\BibitemShut {NoStop}%
\bibitem [{\citenamefont {Xiang}\ \emph {et~al.}(2020)\citenamefont {Xiang}, \citenamefont {Ribeiro}, \citenamefont {Du}, \citenamefont {Chen}, \citenamefont {Yang}, \citenamefont {Wang}, \citenamefont {Yuen-Zhou},\ and\ \citenamefont {Xiong}}]{xiang2020intermolecular}%
  \BibitemOpen
  \bibfield  {author} {\bibinfo {author} {\bibfnamefont {B.}~\bibnamefont {Xiang}}, \bibinfo {author} {\bibfnamefont {R.~F.}\ \bibnamefont {Ribeiro}}, \bibinfo {author} {\bibfnamefont {M.}~\bibnamefont {Du}}, \bibinfo {author} {\bibfnamefont {L.}~\bibnamefont {Chen}}, \bibinfo {author} {\bibfnamefont {Z.}~\bibnamefont {Yang}}, \bibinfo {author} {\bibfnamefont {J.}~\bibnamefont {Wang}}, \bibinfo {author} {\bibfnamefont {J.}~\bibnamefont {Yuen-Zhou}},\ and\ \bibinfo {author} {\bibfnamefont {W.}~\bibnamefont {Xiong}},\ }\bibfield  {title} {\enquote {\bibinfo {title} {Intermolecular vibrational energy transfer enabled by microcavity strong light--matter coupling},}\ }\href@noop {} {\bibfield  {journal} {\bibinfo  {journal} {Science}\ }\textbf {\bibinfo {volume} {368}},\ \bibinfo {pages} {665--667} (\bibinfo {year} {2020})}\BibitemShut {NoStop}%
\bibitem [{\citenamefont {Philbin}\ and\ \citenamefont {Rabani}(2018)}]{philbin2018electron}%
  \BibitemOpen
  \bibfield  {author} {\bibinfo {author} {\bibfnamefont {J.~P.}\ \bibnamefont {Philbin}}\ and\ \bibinfo {author} {\bibfnamefont {E.}~\bibnamefont {Rabani}},\ }\bibfield  {title} {\enquote {\bibinfo {title} {Electron--hole correlations govern auger recombination in nanostructures},}\ }\href@noop {} {\bibfield  {journal} {\bibinfo  {journal} {Nano Lett.}\ }\textbf {\bibinfo {volume} {18}},\ \bibinfo {pages} {7889--7895} (\bibinfo {year} {2018})}\BibitemShut {NoStop}%
\bibitem [{\citenamefont {Jasrasaria}\ and\ \citenamefont {Rabani}(2021)}]{jasrasaria2021interplay}%
  \BibitemOpen
  \bibfield  {author} {\bibinfo {author} {\bibfnamefont {D.}~\bibnamefont {Jasrasaria}}\ and\ \bibinfo {author} {\bibfnamefont {E.}~\bibnamefont {Rabani}},\ }\bibfield  {title} {\enquote {\bibinfo {title} {Interplay of surface and interior modes in exciton--phonon coupling at the nanoscale},}\ }\href@noop {} {\bibfield  {journal} {\bibinfo  {journal} {Nano Lett.}\ }\textbf {\bibinfo {volume} {21}},\ \bibinfo {pages} {8741--8748} (\bibinfo {year} {2021})}\BibitemShut {NoStop}%
\bibitem [{\citenamefont {Jasrasaria}\ \emph {et~al.}(2022)\citenamefont {Jasrasaria}, \citenamefont {Weinberg}, \citenamefont {Philbin},\ and\ \citenamefont {Rabani}}]{jasrasaria2022simulations}%
  \BibitemOpen
  \bibfield  {author} {\bibinfo {author} {\bibfnamefont {D.}~\bibnamefont {Jasrasaria}}, \bibinfo {author} {\bibfnamefont {D.}~\bibnamefont {Weinberg}}, \bibinfo {author} {\bibfnamefont {J.~P.}\ \bibnamefont {Philbin}},\ and\ \bibinfo {author} {\bibfnamefont {E.}~\bibnamefont {Rabani}},\ }\bibfield  {title} {\enquote {\bibinfo {title} {Simulations of nonradiative processes in semiconductor nanocrystals},}\ }\href@noop {} {\bibfield  {journal} {\bibinfo  {journal} {J. Chem. Phys.}\ }\textbf {\bibinfo {volume} {157}},\ \bibinfo {pages} {020901} (\bibinfo {year} {2022})}\BibitemShut {NoStop}%
\bibitem [{\citenamefont {Jasrasaria}\ and\ \citenamefont {Rabani}(2023)}]{jasrasaria2023circumventing}%
  \BibitemOpen
  \bibfield  {author} {\bibinfo {author} {\bibfnamefont {D.}~\bibnamefont {Jasrasaria}}\ and\ \bibinfo {author} {\bibfnamefont {E.}~\bibnamefont {Rabani}},\ }\bibfield  {title} {\enquote {\bibinfo {title} {Circumventing the phonon bottleneck by multiphonon-mediated hot exciton cooling at the nanoscale},}\ }\href@noop {} {\bibfield  {journal} {\bibinfo  {journal} {Npj Comput. Mater.}\ }\textbf {\bibinfo {volume} {9}},\ \bibinfo {pages} {1--8} (\bibinfo {year} {2023})}\BibitemShut {NoStop}%
\bibitem [{\citenamefont {Wall}\ and\ \citenamefont {Neuhauser}(1995)}]{wall1995extraction}%
  \BibitemOpen
  \bibfield  {author} {\bibinfo {author} {\bibfnamefont {M.~R.}\ \bibnamefont {Wall}}\ and\ \bibinfo {author} {\bibfnamefont {D.}~\bibnamefont {Neuhauser}},\ }\bibfield  {title} {\enquote {\bibinfo {title} {Extraction, through filter-diagonalization, of general quantum eigenvalues or classical normal mode frequencies from a small number of residues or a short-time segment of a signal. i. theory and application to a quantum-dynamics model},}\ }\href@noop {} {\bibfield  {journal} {\bibinfo  {journal} {J. Chem. Phys.}\ }\textbf {\bibinfo {volume} {102}},\ \bibinfo {pages} {8011--8022} (\bibinfo {year} {1995})}\BibitemShut {NoStop}%
\bibitem [{\citenamefont {Toledo}\ and\ \citenamefont {Rabani}(2002)}]{toledo2002very}%
  \BibitemOpen
  \bibfield  {author} {\bibinfo {author} {\bibfnamefont {S.}~\bibnamefont {Toledo}}\ and\ \bibinfo {author} {\bibfnamefont {E.}~\bibnamefont {Rabani}},\ }\bibfield  {title} {\enquote {\bibinfo {title} {Very large electronic structure calculations using an out-of-core filter-diagonalization method},}\ }\href@noop {} {\bibfield  {journal} {\bibinfo  {journal} {J. Comput. Phys.}\ }\textbf {\bibinfo {volume} {180}},\ \bibinfo {pages} {256--269} (\bibinfo {year} {2002})}\BibitemShut {NoStop}%
\bibitem [{\citenamefont {Rohlfing}\ and\ \citenamefont {Louie}(2000)}]{rohlfing2000electron}%
  \BibitemOpen
  \bibfield  {author} {\bibinfo {author} {\bibfnamefont {M.}~\bibnamefont {Rohlfing}}\ and\ \bibinfo {author} {\bibfnamefont {S.~G.}\ \bibnamefont {Louie}},\ }\bibfield  {title} {\enquote {\bibinfo {title} {Electron-hole excitations and optical spectra from first principles},}\ }\href@noop {} {\bibfield  {journal} {\bibinfo  {journal} {Phys. Rev. B}\ }\textbf {\bibinfo {volume} {62}},\ \bibinfo {pages} {4927} (\bibinfo {year} {2000})}\BibitemShut {NoStop}%
\bibitem [{\citenamefont {Wang}\ and\ \citenamefont {Zunger}(1996)}]{wang1996pseudopotential}%
  \BibitemOpen
  \bibfield  {author} {\bibinfo {author} {\bibfnamefont {L.-W.}\ \bibnamefont {Wang}}\ and\ \bibinfo {author} {\bibfnamefont {A.}~\bibnamefont {Zunger}},\ }\bibfield  {title} {\enquote {\bibinfo {title} {Pseudopotential calculations of nanoscale cdse quantum dots},}\ }\href@noop {} {\bibfield  {journal} {\bibinfo  {journal} {Phys. Rev. B}\ }\textbf {\bibinfo {volume} {53}},\ \bibinfo {pages} {9579} (\bibinfo {year} {1996})}\BibitemShut {NoStop}%
\bibitem [{\citenamefont {Latt}, \citenamefont {Cheung},\ and\ \citenamefont {Blout}(1965)}]{latt1965energy}%
  \BibitemOpen
  \bibfield  {author} {\bibinfo {author} {\bibfnamefont {S.}~\bibnamefont {Latt}}, \bibinfo {author} {\bibfnamefont {H.}~\bibnamefont {Cheung}},\ and\ \bibinfo {author} {\bibfnamefont {E.}~\bibnamefont {Blout}},\ }\bibfield  {title} {\enquote {\bibinfo {title} {Energy transfer. a system with relatively fixed donor-acceptor separation},}\ }\href@noop {} {\bibfield  {journal} {\bibinfo  {journal} {J. Am. Chem. Soc.}\ }\textbf {\bibinfo {volume} {87}},\ \bibinfo {pages} {995--1003} (\bibinfo {year} {1965})}\BibitemShut {NoStop}%
\bibitem [{\citenamefont {Fedchenia}\ and\ \citenamefont {Westlund}(1994)}]{fedchenia1994influence}%
  \BibitemOpen
  \bibfield  {author} {\bibinfo {author} {\bibfnamefont {I.}~\bibnamefont {Fedchenia}}\ and\ \bibinfo {author} {\bibfnamefont {P.-O.}\ \bibnamefont {Westlund}},\ }\bibfield  {title} {\enquote {\bibinfo {title} {Influence of molecular reorientation on electronic energy transfer between a pair of mobile chromophores: The stochastic liouville equation combined with brownian dynamic simulation techniques},}\ }\href@noop {} {\bibfield  {journal} {\bibinfo  {journal} {Phys. Rev. E.}\ }\textbf {\bibinfo {volume} {50}},\ \bibinfo {pages} {555} (\bibinfo {year} {1994})}\BibitemShut {NoStop}%
\bibitem [{\citenamefont {Zhou}\ \emph {et~al.}(2013)\citenamefont {Zhou}, \citenamefont {Ward}, \citenamefont {Martin}, \citenamefont {Van~Swol}, \citenamefont {Cruz-Campa},\ and\ \citenamefont {Zubia}}]{zhou2013stillinger}%
  \BibitemOpen
  \bibfield  {author} {\bibinfo {author} {\bibfnamefont {X.}~\bibnamefont {Zhou}}, \bibinfo {author} {\bibfnamefont {D.}~\bibnamefont {Ward}}, \bibinfo {author} {\bibfnamefont {J.}~\bibnamefont {Martin}}, \bibinfo {author} {\bibfnamefont {F.}~\bibnamefont {Van~Swol}}, \bibinfo {author} {\bibfnamefont {J.}~\bibnamefont {Cruz-Campa}},\ and\ \bibinfo {author} {\bibfnamefont {D.}~\bibnamefont {Zubia}},\ }\bibfield  {title} {\enquote {\bibinfo {title} {Stillinger-weber potential for the ii-vi elements zn-cd-hg-s-se-te},}\ }\href@noop {} {\bibfield  {journal} {\bibinfo  {journal} {Phys. Rev. B}\ }\textbf {\bibinfo {volume} {88}},\ \bibinfo {pages} {085309} (\bibinfo {year} {2013})}\BibitemShut {NoStop}%
\bibitem [{\citenamefont {Mandal}, \citenamefont {Krauss},\ and\ \citenamefont {Huo}(2020)}]{mandal2020polariton}%
  \BibitemOpen
  \bibfield  {author} {\bibinfo {author} {\bibfnamefont {A.}~\bibnamefont {Mandal}}, \bibinfo {author} {\bibfnamefont {T.~D.}\ \bibnamefont {Krauss}},\ and\ \bibinfo {author} {\bibfnamefont {P.}~\bibnamefont {Huo}},\ }\bibfield  {title} {\enquote {\bibinfo {title} {Polariton-mediated electron transfer via cavity quantum electrodynamics},}\ }\href@noop {} {\bibfield  {journal} {\bibinfo  {journal} {J. Phys. Chem. B}\ }\textbf {\bibinfo {volume} {124}},\ \bibinfo {pages} {6321--6340} (\bibinfo {year} {2020})}\BibitemShut {NoStop}%
\bibitem [{\citenamefont {Peng}\ and\ \citenamefont {Rabani}(2023)}]{peng2023polaritonic}%
  \BibitemOpen
  \bibfield  {author} {\bibinfo {author} {\bibfnamefont {K.}~\bibnamefont {Peng}}\ and\ \bibinfo {author} {\bibfnamefont {E.}~\bibnamefont {Rabani}},\ }\bibfield  {title} {\enquote {\bibinfo {title} {Polaritonic bottleneck in colloidal quantum dots},}\ }\href@noop {} {\bibfield  {journal} {\bibinfo  {journal} {Nano Lett.}\ }\textbf {\bibinfo {volume} {23}},\ \bibinfo {pages} {10587--10593} (\bibinfo {year} {2023})}\BibitemShut {NoStop}%
\bibitem [{\citenamefont {Rokaj}\ \emph {et~al.}(2018)\citenamefont {Rokaj}, \citenamefont {Welakuh}, \citenamefont {Ruggenthaler},\ and\ \citenamefont {Rubio}}]{Rokaj2018}%
  \BibitemOpen
  \bibfield  {author} {\bibinfo {author} {\bibfnamefont {V.}~\bibnamefont {Rokaj}}, \bibinfo {author} {\bibfnamefont {D.~M.}\ \bibnamefont {Welakuh}}, \bibinfo {author} {\bibfnamefont {M.}~\bibnamefont {Ruggenthaler}},\ and\ \bibinfo {author} {\bibfnamefont {A.}~\bibnamefont {Rubio}},\ }\bibfield  {title} {\enquote {\bibinfo {title} {Light−matter interaction in the long-wavelength limit: No ground-statewithout dipole self-energy},}\ }\href@noop {} {\bibfield  {journal} {\bibinfo  {journal} {J. Phys. B: At., Mol. Opt. Phys.}\ }\textbf {\bibinfo {volume} {51}},\ \bibinfo {pages} {034005} (\bibinfo {year} {2018})}\BibitemShut {NoStop}%
\bibitem [{\citenamefont {Sidler}, \citenamefont {Ruggenthaler},\ and\ \citenamefont {Rubio}(2023)}]{sidler2023numerically}%
  \BibitemOpen
  \bibfield  {author} {\bibinfo {author} {\bibfnamefont {D.}~\bibnamefont {Sidler}}, \bibinfo {author} {\bibfnamefont {M.}~\bibnamefont {Ruggenthaler}},\ and\ \bibinfo {author} {\bibfnamefont {A.}~\bibnamefont {Rubio}},\ }\bibfield  {title} {\enquote {\bibinfo {title} {Numerically exact solution for a real polaritonic system under vibrational strong coupling in thermodynamic equilibrium: loss of light--matter entanglement and enhanced fluctuations},}\ }\href@noop {} {\bibfield  {journal} {\bibinfo  {journal} {J. Chem. Theory Comput.}\ }\textbf {\bibinfo {volume} {19}},\ \bibinfo {pages} {8801--8814} (\bibinfo {year} {2023})}\BibitemShut {NoStop}%
\bibitem [{\citenamefont {P{\'e}rez-S{\'a}nchez}\ \emph {et~al.}(2023)\citenamefont {P{\'e}rez-S{\'a}nchez}, \citenamefont {Koner}, \citenamefont {Stern},\ and\ \citenamefont {Yuen-Zhou}}]{perez2023simulating}%
  \BibitemOpen
  \bibfield  {author} {\bibinfo {author} {\bibfnamefont {J.~B.}\ \bibnamefont {P{\'e}rez-S{\'a}nchez}}, \bibinfo {author} {\bibfnamefont {A.}~\bibnamefont {Koner}}, \bibinfo {author} {\bibfnamefont {N.~P.}\ \bibnamefont {Stern}},\ and\ \bibinfo {author} {\bibfnamefont {J.}~\bibnamefont {Yuen-Zhou}},\ }\bibfield  {title} {\enquote {\bibinfo {title} {Simulating molecular polaritons in the collective regime using few-molecule models},}\ }\href@noop {} {\bibfield  {journal} {\bibinfo  {journal} {Proc. Natl. Acad. Sci. U. S. A.}\ }\textbf {\bibinfo {volume} {120}},\ \bibinfo {pages} {e2219223120} (\bibinfo {year} {2023})}\BibitemShut {NoStop}%
\bibitem [{\citenamefont {Nitzan}(2006)}]{nitzan2006chemical}%
  \BibitemOpen
  \bibfield  {author} {\bibinfo {author} {\bibfnamefont {A.}~\bibnamefont {Nitzan}},\ }\href@noop {} {\emph {\bibinfo {title} {Chemical dynamics in condensed phases: relaxation, transfer and reactions in condensed molecular systems}}}\ (\bibinfo  {publisher} {Oxford university press},\ \bibinfo {year} {2006})\BibitemShut {NoStop}%
\bibitem [{\citenamefont {Kagan}\ \emph {et~al.}(1996)\citenamefont {Kagan}, \citenamefont {Murray}, \citenamefont {Nirmal},\ and\ \citenamefont {Bawendi}}]{kagan1996electronic}%
  \BibitemOpen
  \bibfield  {author} {\bibinfo {author} {\bibfnamefont {C.}~\bibnamefont {Kagan}}, \bibinfo {author} {\bibfnamefont {C.}~\bibnamefont {Murray}}, \bibinfo {author} {\bibfnamefont {M.}~\bibnamefont {Nirmal}},\ and\ \bibinfo {author} {\bibfnamefont {M.}~\bibnamefont {Bawendi}},\ }\bibfield  {title} {\enquote {\bibinfo {title} {Electronic energy transfer in cdse quantum dot solids},}\ }\href@noop {} {\bibfield  {journal} {\bibinfo  {journal} {Phys. Rev. Lett.}\ }\textbf {\bibinfo {volume} {76}},\ \bibinfo {pages} {1517} (\bibinfo {year} {1996})}\BibitemShut {NoStop}%
\bibitem [{\citenamefont {Brosseau}\ \emph {et~al.}(2023)\citenamefont {Brosseau}, \citenamefont {Geuchies}, \citenamefont {Jasrasaria}, \citenamefont {Houtepen}, \citenamefont {Rabani},\ and\ \citenamefont {Kambhampati}}]{brosseau2023ultrafast}%
  \BibitemOpen
  \bibfield  {author} {\bibinfo {author} {\bibfnamefont {P.~J.}\ \bibnamefont {Brosseau}}, \bibinfo {author} {\bibfnamefont {J.~J.}\ \bibnamefont {Geuchies}}, \bibinfo {author} {\bibfnamefont {D.}~\bibnamefont {Jasrasaria}}, \bibinfo {author} {\bibfnamefont {A.~J.}\ \bibnamefont {Houtepen}}, \bibinfo {author} {\bibfnamefont {E.}~\bibnamefont {Rabani}},\ and\ \bibinfo {author} {\bibfnamefont {P.}~\bibnamefont {Kambhampati}},\ }\bibfield  {title} {\enquote {\bibinfo {title} {Ultrafast hole relaxation dynamics in quantum dots revealed by two-dimensional electronic spectroscopy},}\ }\href@noop {} {\bibfield  {journal} {\bibinfo  {journal} {Commun. Phys.}\ }\textbf {\bibinfo {volume} {6}},\ \bibinfo {pages} {48} (\bibinfo {year} {2023})}\BibitemShut {NoStop}%
\bibitem [{\citenamefont {Rabouw}\ \emph {et~al.}(2015)\citenamefont {Rabouw}, \citenamefont {Kamp}, \citenamefont {van Dijk-Moes}, \citenamefont {Gamelin}, \citenamefont {Koenderink}, \citenamefont {Meijerink},\ and\ \citenamefont {Vanmaekelbergh}}]{rabouw2015delayed}%
  \BibitemOpen
  \bibfield  {author} {\bibinfo {author} {\bibfnamefont {F.~T.}\ \bibnamefont {Rabouw}}, \bibinfo {author} {\bibfnamefont {M.}~\bibnamefont {Kamp}}, \bibinfo {author} {\bibfnamefont {R.~J.}\ \bibnamefont {van Dijk-Moes}}, \bibinfo {author} {\bibfnamefont {D.~R.}\ \bibnamefont {Gamelin}}, \bibinfo {author} {\bibfnamefont {A.~F.}\ \bibnamefont {Koenderink}}, \bibinfo {author} {\bibfnamefont {A.}~\bibnamefont {Meijerink}},\ and\ \bibinfo {author} {\bibfnamefont {D.}~\bibnamefont {Vanmaekelbergh}},\ }\bibfield  {title} {\enquote {\bibinfo {title} {Delayed exciton emission and its relation to blinking in cdse quantum dots},}\ }\href@noop {} {\bibfield  {journal} {\bibinfo  {journal} {Nano Lett.}\ }\textbf {\bibinfo {volume} {15}},\ \bibinfo {pages} {7718--7725} (\bibinfo {year} {2015})}\BibitemShut {NoStop}%
\bibitem [{\citenamefont {Bae}\ \emph {et~al.}(2013{\natexlab{a}})\citenamefont {Bae}, \citenamefont {Padilha}, \citenamefont {Park}, \citenamefont {McDaniel}, \citenamefont {Robel}, \citenamefont {Pietryga},\ and\ \citenamefont {Klimov}}]{bae2013controlled}%
  \BibitemOpen
  \bibfield  {author} {\bibinfo {author} {\bibfnamefont {W.~K.}\ \bibnamefont {Bae}}, \bibinfo {author} {\bibfnamefont {L.~A.}\ \bibnamefont {Padilha}}, \bibinfo {author} {\bibfnamefont {Y.-S.}\ \bibnamefont {Park}}, \bibinfo {author} {\bibfnamefont {H.}~\bibnamefont {McDaniel}}, \bibinfo {author} {\bibfnamefont {I.}~\bibnamefont {Robel}}, \bibinfo {author} {\bibfnamefont {J.~M.}\ \bibnamefont {Pietryga}},\ and\ \bibinfo {author} {\bibfnamefont {V.~I.}\ \bibnamefont {Klimov}},\ }\bibfield  {title} {\enquote {\bibinfo {title} {Controlled alloying of the core--shell interface in cdse/cds quantum dots for suppression of auger recombination},}\ }\href@noop {} {\bibfield  {journal} {\bibinfo  {journal} {ACS nano}\ }\textbf {\bibinfo {volume} {7}},\ \bibinfo {pages} {3411--3419} (\bibinfo {year} {2013}{\natexlab{a}})}\BibitemShut {NoStop}%
\bibitem [{\citenamefont {Bae}\ \emph {et~al.}(2013{\natexlab{b}})\citenamefont {Bae}, \citenamefont {Park}, \citenamefont {Lim}, \citenamefont {Lee}, \citenamefont {Padilha}, \citenamefont {McDaniel}, \citenamefont {Robel}, \citenamefont {Lee}, \citenamefont {Pietryga},\ and\ \citenamefont {Klimov}}]{bae2013controlling}%
  \BibitemOpen
  \bibfield  {author} {\bibinfo {author} {\bibfnamefont {W.~K.}\ \bibnamefont {Bae}}, \bibinfo {author} {\bibfnamefont {Y.-S.}\ \bibnamefont {Park}}, \bibinfo {author} {\bibfnamefont {J.}~\bibnamefont {Lim}}, \bibinfo {author} {\bibfnamefont {D.}~\bibnamefont {Lee}}, \bibinfo {author} {\bibfnamefont {L.~A.}\ \bibnamefont {Padilha}}, \bibinfo {author} {\bibfnamefont {H.}~\bibnamefont {McDaniel}}, \bibinfo {author} {\bibfnamefont {I.}~\bibnamefont {Robel}}, \bibinfo {author} {\bibfnamefont {C.}~\bibnamefont {Lee}}, \bibinfo {author} {\bibfnamefont {J.~M.}\ \bibnamefont {Pietryga}},\ and\ \bibinfo {author} {\bibfnamefont {V.~I.}\ \bibnamefont {Klimov}},\ }\bibfield  {title} {\enquote {\bibinfo {title} {Controlling the influence of auger recombination on the performance of quantum-dot light-emitting diodes},}\ }\href@noop {} {\bibfield  {journal} {\bibinfo  {journal} {Nat. Commun.}\ }\textbf {\bibinfo {volume} {4}},\ \bibinfo {pages} {2661} (\bibinfo {year} {2013}{\natexlab{b}})}\BibitemShut {NoStop}%
\bibitem [{\citenamefont {Crooker}\ \emph {et~al.}(2002)\citenamefont {Crooker}, \citenamefont {Hollingsworth}, \citenamefont {Tretiak},\ and\ \citenamefont {Klimov}}]{crooker2002spectrally}%
  \BibitemOpen
  \bibfield  {author} {\bibinfo {author} {\bibfnamefont {S.}~\bibnamefont {Crooker}}, \bibinfo {author} {\bibfnamefont {J.}~\bibnamefont {Hollingsworth}}, \bibinfo {author} {\bibfnamefont {S.}~\bibnamefont {Tretiak}},\ and\ \bibinfo {author} {\bibfnamefont {V.~I.}\ \bibnamefont {Klimov}},\ }\bibfield  {title} {\enquote {\bibinfo {title} {Spectrally resolved dynamics of energy transfer in quantum-dot assemblies: towards engineered energy flows in artificial materials},}\ }\href@noop {} {\bibfield  {journal} {\bibinfo  {journal} {Phys. Rev. Lett.}\ }\textbf {\bibinfo {volume} {89}},\ \bibinfo {pages} {186802} (\bibinfo {year} {2002})}\BibitemShut {NoStop}%
\bibitem [{\citenamefont {Lin}\ \emph {et~al.}(2023)\citenamefont {Lin}, \citenamefont {Jasrasaria}, \citenamefont {Yoo}, \citenamefont {Bawendi}, \citenamefont {Utzat},\ and\ \citenamefont {Rabani}}]{kailai2023lumi}%
  \BibitemOpen
  \bibfield  {author} {\bibinfo {author} {\bibfnamefont {K.}~\bibnamefont {Lin}}, \bibinfo {author} {\bibfnamefont {D.}~\bibnamefont {Jasrasaria}}, \bibinfo {author} {\bibfnamefont {J.~J.}\ \bibnamefont {Yoo}}, \bibinfo {author} {\bibfnamefont {M.}~\bibnamefont {Bawendi}}, \bibinfo {author} {\bibfnamefont {H.}~\bibnamefont {Utzat}},\ and\ \bibinfo {author} {\bibfnamefont {E.}~\bibnamefont {Rabani}},\ }\bibfield  {title} {\enquote {\bibinfo {title} {Theory of photoluminescence spectral line shapes of semiconductor nanocrystals},}\ }\href@noop {} {\bibfield  {journal} {\bibinfo  {journal} {J. Phys. Chem. Lett.}\ }\textbf {\bibinfo {volume} {14}},\ \bibinfo {pages} {7241--7248} (\bibinfo {year} {2023})}\BibitemShut {NoStop}%
\bibitem [{\citenamefont {Lingley}, \citenamefont {Lu},\ and\ \citenamefont {Madhukar}(2011)}]{lingley2011high}%
  \BibitemOpen
  \bibfield  {author} {\bibinfo {author} {\bibfnamefont {Z.}~\bibnamefont {Lingley}}, \bibinfo {author} {\bibfnamefont {S.}~\bibnamefont {Lu}},\ and\ \bibinfo {author} {\bibfnamefont {A.}~\bibnamefont {Madhukar}},\ }\bibfield  {title} {\enquote {\bibinfo {title} {A high quantum efficiency preserving approach to ligand exchange on lead sulfide quantum dots and interdot resonant energy transfer},}\ }\href@noop {} {\bibfield  {journal} {\bibinfo  {journal} {Nano Lett.}\ }\textbf {\bibinfo {volume} {11}},\ \bibinfo {pages} {2887--2891} (\bibinfo {year} {2011})}\BibitemShut {NoStop}%
\bibitem [{\citenamefont {le~Feber}\ \emph {et~al.}(2018)\citenamefont {le~Feber}, \citenamefont {Prins}, \citenamefont {De~Leo}, \citenamefont {Rabouw},\ and\ \citenamefont {Norris}}]{le2018colloidal}%
  \BibitemOpen
  \bibfield  {author} {\bibinfo {author} {\bibfnamefont {B.}~\bibnamefont {le~Feber}}, \bibinfo {author} {\bibfnamefont {F.}~\bibnamefont {Prins}}, \bibinfo {author} {\bibfnamefont {E.}~\bibnamefont {De~Leo}}, \bibinfo {author} {\bibfnamefont {F.~T.}\ \bibnamefont {Rabouw}},\ and\ \bibinfo {author} {\bibfnamefont {D.~J.}\ \bibnamefont {Norris}},\ }\bibfield  {title} {\enquote {\bibinfo {title} {Colloidal-quantum-dot ring lasers with active color control},}\ }\href@noop {} {\bibfield  {journal} {\bibinfo  {journal} {Nano Lett.}\ }\textbf {\bibinfo {volume} {18}},\ \bibinfo {pages} {1028--1034} (\bibinfo {year} {2018})}\BibitemShut {NoStop}%
\bibitem [{\citenamefont {Davis}\ \emph {et~al.}(2009)\citenamefont {Davis}, \citenamefont {Toroker}, \citenamefont {Speiser},\ and\ \citenamefont {Peskin}}]{davis2009effect}%
  \BibitemOpen
  \bibfield  {author} {\bibinfo {author} {\bibfnamefont {D.}~\bibnamefont {Davis}}, \bibinfo {author} {\bibfnamefont {M.~C.}\ \bibnamefont {Toroker}}, \bibinfo {author} {\bibfnamefont {S.}~\bibnamefont {Speiser}},\ and\ \bibinfo {author} {\bibfnamefont {U.}~\bibnamefont {Peskin}},\ }\bibfield  {title} {\enquote {\bibinfo {title} {On the effect of nuclear bridge modes on donor--acceptor electronic coupling in donor--bridge--acceptor molecules},}\ }\href@noop {} {\bibfield  {journal} {\bibinfo  {journal} {Chem. Phys.}\ }\textbf {\bibinfo {volume} {358}},\ \bibinfo {pages} {45--51} (\bibinfo {year} {2009})}\BibitemShut {NoStop}%
\bibitem [{\citenamefont {Skourtis}, \citenamefont {Archontis},\ and\ \citenamefont {Xie}(2001)}]{skourtis2001electron}%
  \BibitemOpen
  \bibfield  {author} {\bibinfo {author} {\bibfnamefont {S.~S.}\ \bibnamefont {Skourtis}}, \bibinfo {author} {\bibfnamefont {G.}~\bibnamefont {Archontis}},\ and\ \bibinfo {author} {\bibfnamefont {Q.}~\bibnamefont {Xie}},\ }\bibfield  {title} {\enquote {\bibinfo {title} {Electron transfer through fluctuating bridges: On the validity of the superexchange mechanism and time-dependent tunneling matrix elements},}\ }\href@noop {} {\bibfield  {journal} {\bibinfo  {journal} {J. Chem. Phys.}\ }\textbf {\bibinfo {volume} {115}},\ \bibinfo {pages} {9444--9462} (\bibinfo {year} {2001})}\BibitemShut {NoStop}%
\bibitem [{\citenamefont {Priyadarshy}\ \emph {et~al.}(1996)\citenamefont {Priyadarshy}, \citenamefont {Skourtis}, \citenamefont {Risser},\ and\ \citenamefont {Beratan}}]{priyadarshy1996bridge}%
  \BibitemOpen
  \bibfield  {author} {\bibinfo {author} {\bibfnamefont {S.}~\bibnamefont {Priyadarshy}}, \bibinfo {author} {\bibfnamefont {S.~S.}\ \bibnamefont {Skourtis}}, \bibinfo {author} {\bibfnamefont {S.~M.}\ \bibnamefont {Risser}},\ and\ \bibinfo {author} {\bibfnamefont {D.~N.}\ \bibnamefont {Beratan}},\ }\bibfield  {title} {\enquote {\bibinfo {title} {Bridge-mediated electronic interactions: Differences between hamiltonian and green function partitioning in a non-orthogonal basis},}\ }\href@noop {} {\bibfield  {journal} {\bibinfo  {journal} {J. Chem. Phys.}\ }\textbf {\bibinfo {volume} {104}},\ \bibinfo {pages} {9473--9481} (\bibinfo {year} {1996})}\BibitemShut {NoStop}%
\bibitem [{\citenamefont {Hou}\ \emph {et~al.}(2023)\citenamefont {Hou}, \citenamefont {Thoss}, \citenamefont {Banin},\ and\ \citenamefont {Rabani}}]{hou_incoherent_2023}%
  \BibitemOpen
  \bibfield  {author} {\bibinfo {author} {\bibfnamefont {B.}~\bibnamefont {Hou}}, \bibinfo {author} {\bibfnamefont {M.}~\bibnamefont {Thoss}}, \bibinfo {author} {\bibfnamefont {U.}~\bibnamefont {Banin}},\ and\ \bibinfo {author} {\bibfnamefont {E.}~\bibnamefont {Rabani}},\ }\bibfield  {title} {\enquote {\bibinfo {title} {Incoherent nonadiabatic to coherent adiabatic transition of electron transfer in colloidal quantum dot molecules},}\ }\href@noop {} {\bibfield  {journal} {\bibinfo  {journal} {Nat. Commun.}\ }\textbf {\bibinfo {volume} {14}},\ \bibinfo {pages} {3073} (\bibinfo {year} {2023})}\BibitemShut {NoStop}%
\bibitem [{\citenamefont {Newton}\ and\ \citenamefont {Sutin}(1984)}]{newton1984electron}%
  \BibitemOpen
  \bibfield  {author} {\bibinfo {author} {\bibfnamefont {M.~D.}\ \bibnamefont {Newton}}\ and\ \bibinfo {author} {\bibfnamefont {N.}~\bibnamefont {Sutin}},\ }\bibfield  {title} {\enquote {\bibinfo {title} {Electron transfer reactions in condensed phases},}\ }\href@noop {} {\bibfield  {journal} {\bibinfo  {journal} {Annu. Rev. Phys. Chem.}\ }\textbf {\bibinfo {volume} {35}},\ \bibinfo {pages} {437--480} (\bibinfo {year} {1984})}\BibitemShut {NoStop}%
\bibitem [{\citenamefont {Balzani}\ \emph {et~al.}(2001)\citenamefont {Balzani}, \citenamefont {Piotrowiak}, \citenamefont {Rodgers}, \citenamefont {Mattay}, \citenamefont {Astruc} \emph {et~al.}}]{balzani2001electron}%
  \BibitemOpen
  \bibfield  {author} {\bibinfo {author} {\bibfnamefont {V.}~\bibnamefont {Balzani}}, \bibinfo {author} {\bibfnamefont {P.}~\bibnamefont {Piotrowiak}}, \bibinfo {author} {\bibfnamefont {M.}~\bibnamefont {Rodgers}}, \bibinfo {author} {\bibfnamefont {J.}~\bibnamefont {Mattay}}, \bibinfo {author} {\bibfnamefont {D.}~\bibnamefont {Astruc}}, \emph {et~al.},\ }\href@noop {} {\emph {\bibinfo {title} {Electron transfer in chemistry}}},\ Vol.~\bibinfo {volume} {1}\ (\bibinfo  {publisher} {Wiley-VCh Weinheim},\ \bibinfo {year} {2001})\BibitemShut {NoStop}%
\bibitem [{\citenamefont {Chikkaraddy}\ \emph {et~al.}(2016)\citenamefont {Chikkaraddy}, \citenamefont {De~Nijs}, \citenamefont {Benz}, \citenamefont {Barrow}, \citenamefont {Scherman}, \citenamefont {Rosta}, \citenamefont {Demetriadou}, \citenamefont {Fox}, \citenamefont {Hess},\ and\ \citenamefont {Baumberg}}]{chikkaraddy2016single}%
  \BibitemOpen
  \bibfield  {author} {\bibinfo {author} {\bibfnamefont {R.}~\bibnamefont {Chikkaraddy}}, \bibinfo {author} {\bibfnamefont {B.}~\bibnamefont {De~Nijs}}, \bibinfo {author} {\bibfnamefont {F.}~\bibnamefont {Benz}}, \bibinfo {author} {\bibfnamefont {S.~J.}\ \bibnamefont {Barrow}}, \bibinfo {author} {\bibfnamefont {O.~A.}\ \bibnamefont {Scherman}}, \bibinfo {author} {\bibfnamefont {E.}~\bibnamefont {Rosta}}, \bibinfo {author} {\bibfnamefont {A.}~\bibnamefont {Demetriadou}}, \bibinfo {author} {\bibfnamefont {P.}~\bibnamefont {Fox}}, \bibinfo {author} {\bibfnamefont {O.}~\bibnamefont {Hess}},\ and\ \bibinfo {author} {\bibfnamefont {J.~J.}\ \bibnamefont {Baumberg}},\ }\bibfield  {title} {\enquote {\bibinfo {title} {Single-molecule strong coupling at room temperature in plasmonic nanocavities},}\ }\href@noop {} {\bibfield  {journal} {\bibinfo  {journal} {Nature}\ }\textbf {\bibinfo {volume} {535}},\ \bibinfo {pages} {127--130} (\bibinfo {year} {2016})}\BibitemShut {NoStop}%
\end{thebibliography}%
\end{document}


\title{Supporting Information: Polariton assisted incoherent to coherent excitation energy transfer between colloidal quantum dots}

\author{Kaiyue Peng}
\email{kaiyue\_peng@berkeley.edu}
\affiliation{Department of Chemistry, University of California, Berkeley, California
94720, United States}

\author{Eran Rabani}
\email{eran.rabani@berkeley.edu}
\affiliation{Department of Chemistry, University of California, Berkeley, California
94720, United States}
\affiliation{Materials Sciences Division, Lawrence Berkeley National Laboratory,
Berkeley, California 94720, United States}
\affiliation{The Sackler Center for Computational Molecular and Materials Science,
Tel Aviv University, Tel Aviv, Israel 69978}

\maketitle
\section{Nanocrystal Configurations}
We consider $3$ quantum dot (QD) systems. The $3$ nm CdSe core with a $2$-monolayer
CdS shell QD contains a total of $753$ Cd atoms, $252$
Se atoms, and $501$ S atoms. The $3.9$ nm CdSe core with a $3$-monolayer
CdS shell QD contains a total of $1788$ Cd atoms, $483$ Se atoms,
and $1305$ S atoms. The $3.9$ nm CdSe core with a $4$-monolayer
CdS shell QD contains a total of $2637$ Cd atoms, $483$ Se atoms,
and $2154$ S atoms.

The Stillinger--Weber force field~\cite{zhou2013stillinger} was
utilized to determine the equilibrium structure and normal vibrational
modes using LAMMPS.\cite{plimpton1995fast} Following minimization,
the outermost layer was replaced by ligands with a modification of
the pseudopotential to represent the passivation layer.\cite{rabani1999electronic}

\section{Polaron transformation}
The total Hamiltonian in polaritonic basis:
\begin{align}\tilde{H} & =\sum_{n}\tilde{E}_{n}\left|\varphi_{n}\right\rangle \left\langle \varphi_{n}\right|+\sum_{\alpha}\hbar\omega_{\alpha}b_{\alpha}^{\dagger}b_{\alpha}+\sum_{\alpha nm}\tilde{V}_{nm}^{\alpha}\left|\varphi_{n}\right\rangle \left\langle \varphi_{m}\right|q_{\alpha}.\end{align}
Apply a unitary polaron transformation:\citep{jasrasaria2023circumventing,peng2023polaritonic}
\begin{align}
{\cal H} & =e^{S}\tilde{H}e^{-S}=\exp\left(-\sum_{\alpha}\frac{i\sum_{n}\tilde{V}_{nn}^{\alpha}p_{\alpha}}{\hbar\omega_{\alpha}^{2}}\left|\varphi_{n}\right\rangle \left\langle \varphi_{n}\right|\right)\tilde{H}\exp\left(\sum_{\alpha}\frac{i\sum_{n}\tilde{V}_{nn}^{\alpha}p_{\alpha}}{\hbar\omega_{\alpha}^{2}}\left|\varphi_{n}\right\rangle \left\langle \varphi_{n}\right|\right)\\
& =\exp\left(-\sum_{\alpha}\frac{i\sum_{n}\tilde{V}_{nn}^{\alpha}p_{\alpha}}{\hbar\omega_{\alpha}^{2}}\left|\varphi_{n}\right\rangle \left\langle \varphi_{n}\right|\right)\left[\sum_{\alpha}\hbar\omega_{\alpha}b_{\alpha}^{\dagger}b_{\alpha}\right]\exp\left(\sum_{\alpha}\frac{i\sum_{n}\tilde{V}_{nn}^{\alpha}p_{\alpha}}{\hbar\omega_{\alpha}^{2}}\left|\varphi_{n}\right\rangle \left\langle \varphi_{n}\right|\right)\label{eq:PT_bath}\\
& +\exp\left(-\sum_{\alpha}\frac{i\sum_{n}\tilde{V}_{nn}^{\alpha}p_{\alpha}}{\hbar\omega_{\alpha}^{2}}\left|\varphi_{n}\right\rangle \left\langle \varphi_{n}\right|\right)\left[\sum_{n}\tilde{E}_{n}\left|\varphi_{n}\right\rangle \left\langle \varphi_{n}\right|\right]\exp\left(\sum_{\alpha}\frac{i\sum_{n}\tilde{V}_{nn}^{\alpha}p_{\alpha}}{\hbar\omega_{\alpha}^{2}}\left|\varphi_{n}\right\rangle \left\langle \varphi_{n}\right|\right)\label{eq:PT_diagsys}\\
& +\exp\left(-\sum_{\alpha}\frac{i\sum_{n}\tilde{V}_{nn}^{\alpha}p_{\alpha}}{\hbar\omega_{\alpha}^{2}}\left|\varphi_{n}\right\rangle \left\langle \varphi_{n}\right|\right)\left[\sum_{\alpha nm}\tilde{V}_{nm}^{\alpha}\left|\varphi_{n}\right\rangle \left\langle \varphi_{m}\right|q_{\alpha}\right]\exp\left(\sum_{\alpha}\frac{i\sum_{n}\tilde{V}_{nn}^{\alpha}p_{\alpha}}{\hbar\omega_{\alpha}^{2}}\left|\varphi_{n}\right\rangle \left\langle \varphi_{n}\right|\right).\label{eq:PT_EXPC}
\end{align}
Use the relationship:
\begin{align}
U(\lambda) & =e^{-\lambda\frac{\partial}{\partial q}}=e^{-\lambda\frac{ip}{\hbar}},\\
U(\lambda)qU^{\dagger}(\lambda) & =e^{-\lambda\frac{ip}{\hbar}}qe^{\lambda\frac{ip}{\hbar}}=q-\lambda,\\
U(\lambda)pU^{\dagger}(\lambda) & =e^{-\lambda\frac{ip}{\hbar}}pe^{\lambda\frac{ip}{\hbar}}=p.
\end{align}
Eq. \ref{eq:PT_bath} transforms the phonon bath part:
\begin{align}
& \exp\left(-\sum_{\alpha}\frac{i\sum_{n}\tilde{V}_{nn}^{\alpha}p_{\alpha}}{\hbar\omega_{\alpha}^{2}}\left|\varphi_{n}\right\rangle \left\langle \varphi_{n}\right|\right)\left[\sum_{\alpha}\hbar\omega_{\alpha}b_{\alpha}^{\dagger}b_{\alpha}\right]\exp\left(\sum_{\alpha}\frac{i\sum_{n}\tilde{V}_{nn}^{\alpha}p_{\alpha}}{\hbar\omega_{\alpha}^{2}}\left|\varphi_{n}\right\rangle \left\langle \varphi_{n}\right|\right)\nonumber \\
& =\exp\left(-\sum_{\alpha}\frac{i\sum_{n}\tilde{V}_{nn}^{\alpha}p_{\alpha}}{\hbar\omega_{\alpha}^{2}}\left|\varphi_{n}\right\rangle \left\langle \varphi_{n}\right|\right)\left[\sum_{\alpha}\left(\frac{p_{\alpha}^{2}}{2m}+\frac{1}{2}\omega_{\alpha}^{2}q_{\alpha}^{2}\right)\right]\exp\left(\sum_{\alpha}\frac{i\sum_{n}\tilde{V}_{nn}^{\alpha}p_{\alpha}}{\hbar\omega_{\alpha}^{2}}\left|\varphi_{n}\right\rangle \left\langle \varphi_{n}\right|\right)\nonumber \\
& =\sum_{\alpha}\left[\frac{p_{\alpha}^{2}}{2m}+\frac{1}{2}\omega_{\alpha}^{2}\left(q_{\alpha}-\frac{\sum_{n}\left(\tilde{V}_{nn}^{\alpha}\left|\varphi_{n}\right\rangle \left\langle \varphi_{n}\right|\right)}{\omega_{\alpha}^{2}}\right)^{2}\right]\nonumber \\
& =\sum_{\alpha}\hbar\omega_{\alpha}b_{\alpha}^{\dagger}b_{\alpha}+\sum_{n}\left[\sum_{\alpha}\frac{\left(\tilde{V}_{nn}^{\alpha}\right)^{2}}{2\omega_{\alpha}^{2}}\left|\varphi_{n}\right\rangle \left\langle \varphi_{n}\right|\right]-\sum_{n}\left(\sum_{\alpha}\tilde{V}_{nn}^{\alpha}q_{\alpha}\left|\varphi_{n}\right\rangle \left\langle \varphi_{n}\right|\right).
\end{align}
Eq. \ref{eq:PT_diagsys} transforms the diagonal system part, which is unchanged:
\begin{align}
& \exp\left(-\sum_{\alpha}\frac{i\sum_{n}\tilde{V}_{nn}^{\alpha}p_{\alpha}}{\hbar\omega_{\alpha}^{2}}\left|\varphi_{n}\right\rangle \left\langle \varphi_{n}\right|\right)\left[\sum_{n}\tilde{E}_{n}\left|\varphi_{n}\right\rangle \left\langle \varphi_{n}\right|\right]\exp\left(\sum_{\alpha}\frac{i\sum_{n}\tilde{V}_{nn}^{\alpha}p_{\alpha}}{\hbar\omega_{\alpha}^{2}}\left|\varphi_{n}\right\rangle \left\langle \varphi_{n}\right|\right)\nonumber \\
& =\sum_{n}\tilde{E}_{n}\left|\varphi_{n}\right\rangle \left\langle \varphi_{n}\right|.
\end{align}
Eq. \ref{eq:PT_EXPC} transforms the polariton-phonon interaction part:
\begin{align}
& \exp\left(-\sum_{\alpha}\frac{i\sum_{n}\tilde{V}_{nn}^{\alpha}p_{\alpha}}{\hbar\omega_{\alpha}^{2}}\left|\varphi_{n}\right\rangle \left\langle \varphi_{n}\right|\right)\left[\sum_{\alpha nm}\tilde{V}_{nm}^{\alpha}\left|\varphi_{n}\right\rangle \left\langle \varphi_{m}\right|q_{\alpha}\right]\exp\left(\sum_{\alpha}\frac{i\sum_{n}\tilde{V}_{nn}^{\alpha}p_{\alpha}}{\hbar\omega_{\alpha}^{2}}\left|\varphi_{n}\right\rangle \left\langle \varphi_{n}\right|\right)\nonumber \\
& =\sum_{nm}\left[\exp\left(-\sum_{\alpha}\frac{i\tilde{V}_{nn}^{\alpha}p_{\alpha}}{\hbar\omega_{\alpha}^{2}}\right)\left[\sum_{\alpha}\tilde{V}_{nm}^{\alpha}q_{\alpha}\right]\exp\left(\sum_{\alpha}\frac{i\tilde{V}_{mm}^{\alpha}p_{\alpha}}{\hbar\omega_{\alpha}^{2}}\right)\right]\left|\varphi_{n}\right\rangle \left\langle \varphi_{m}\right|\nonumber \\
& =\sum_{n}\left(\sum_{\alpha}\tilde{V}_{nn}^{\alpha}q_{\alpha}-\sum_{\alpha}\frac{\left(\tilde{V}_{nn}^{\alpha}\right)^{2}}{\omega_{\alpha}^{2}}\right)\left|\varphi_{n}\right\rangle \left\langle \varphi_{n}\right|\nonumber \\
& +\sum_{n\neq m}\exp\left(-\sum_{\alpha}\frac{i\tilde{V}_{nn}^{\alpha}p_{\alpha}}{\hbar\omega_{\alpha}^{2}}\right)\left[\sum_{\alpha}\tilde{V}_{nm}^{\alpha}q_{\alpha}\right]\exp\left(\sum_{\alpha}\frac{i\tilde{V}_{mm}^{\alpha}p_{\alpha}}{\hbar\omega_{\alpha}^{2}}\right)\left|\varphi_{n}\right\rangle \left\langle \varphi_{m}\right|.
\end{align}
Then the transformed total Hamiltonian is given by:
\begin{align}
{\cal H} & =\sum_{n}\left(\tilde{E}_{n}-\sum_{\alpha}\frac{\left(\tilde{V}_{nn}^{\alpha}\right)^{2}}{2\omega_{\alpha}^{2}}\right)\left|\varphi_{n}\right\rangle \left\langle \varphi_{n}\right|+\sum_{\alpha}\hbar\omega_{\alpha}b_{\alpha}^{\dagger}b_{\alpha}\nonumber \\
& +\sum_{n\neq m}\exp\left(-\sum_{\alpha}\frac{i\tilde{V}_{nn}^{\alpha}p_{\alpha}}{\hbar\omega_{\alpha}^{2}}\right)\left[\sum_{\alpha}\tilde{V}_{nm}^{\alpha}q_{\alpha}\right]\exp\left(\sum_{\alpha}\frac{i\tilde{V}_{mm}^{\alpha}p_{\alpha}}{\hbar\omega_{\alpha}^{2}}\right)\left|\varphi_{n}\right\rangle \left\langle \varphi_{m}\right|.
\end{align}
We label reorganization energy $\lambda_{n}=\frac{1}{2}\sum_{\alpha}\left(\tilde{V}_{nn}^{\alpha}\right)^{2}/\omega_{\alpha}^{2}$
and the transformed polariton-phonon coupling $W_{nm}=\exp\left(-\sum_{\alpha}\frac{i\tilde{V}_{nn}^{\alpha}p_{\alpha}}{\hbar\omega_{\alpha}^{2}}\right)\left[\sum_{\alpha}\tilde{V}_{nm}^{\alpha}q_{\alpha}\right]\exp\left(\sum_{\alpha}\frac{i\tilde{V}_{mm}^{\alpha}p_{\alpha}}{\hbar\omega_{\alpha}^{2}}\right)$,
then total Hamiltonian $\mathcal{H}$ can be written as:
\begin{align}
{\cal H} & =\overset{{\scriptstyle \mathcal{H}_{{\rm S}}}}{\overbrace{\sum_{n}\left(\tilde{E}_{n}-\lambda_{n}\right)\left|\varphi_{n}\right\rangle \left\langle \varphi_{n}\right|}}+\overset{{\scriptstyle \mathcal{H}_{\text{B}}}}{\overbrace{\sum_{\alpha}\hbar\omega_{\alpha}b_{\alpha}^{\dagger}b_{\alpha}}+}\overset{{\scriptstyle \mathcal{H}_{\text{I}}}}{\overbrace{\sum_{n\neq m}W_{nm}\left|\varphi_{n}\right\rangle \left\langle \varphi_{m}\right|}}.
\end{align}

\section{Time--local Redfield Equations}
We use polaritonic basis $\left| \varphi_{n}\right\rangle$, to express the reduced density matrix and the system Hamiltonian:
\begin{align}
&\sigma_{nm}\left(t\right)=\left\langle \varphi_{n}\right|\sigma\left(t\right)\left|\varphi_{m}\right\rangle,\\
&\left\langle \varphi_{n}\right|\left[\mathcal{H}_{\text{S}},\sigma\left(t\right)\right]\left|\varphi_{m}\right\rangle /\hbar=\omega_{nm}\sigma_{nm}.
\end{align}
Using second order perturbation in $\mathcal{H}_{\text{I}}$, the equation of motion of the reduced density matrix is given by:\citep{nitzan2006chemical}
\begin{align}
\frac{\partial\sigma_{nm}}{\partial t} & =-i\omega_{nm}\sigma_{nm}-\frac{i}{\hbar}\left[\sum_{l\neq n}\left\langle W_{nl}\right\rangle _{{\rm B}}\sigma_{lm}-\sum_{l\neq m}\sigma_{nl}\left\langle W_{lm}\right\rangle _{{\rm B}}\right]\nonumber \\
 & +\sum_{kl}\int_{0}^{t}d\tau\left[-M_{nk,kl}\left(\tau\right)e^{i\omega_{mk}\tau}\sigma_{lm}\left(t-\tau\right)-M_{kl,lm}\left(-\tau\right)e^{i\omega_{ln}\tau}\sigma_{nk}\left(t-\tau\right)\right.\nonumber \\
 & \left.+M_{nk,lm}\left(\tau\right)e^{i\omega_{ln}\tau}\sigma_{kl}\left(t-\tau\right)+M_{nk,lm}\left(-\tau\right)e^{i\omega_{mk}\tau}\sigma_{kl}\left(t-\tau\right)\right],
\end{align}
where 
\begin{align}
\text{if }n\neq m\text{ and }k\neq l:~ & M_{nm,kl}\left(\tau\right)=\frac{1}{\hbar^{2}}\left[\left\langle W_{nm}\left(\tau\right)W_{kl}(0)\right\rangle _{{\rm B}}-\left\langle W_{nm}\right\rangle _{{\rm B}}\left\langle W_{kl}\right\rangle _{{\rm B}}\right],\\
\text{else:}~ & M_{nm,kl}\left(\tau\right)=0.
\end{align}
Employing a time local approximation, with $e^{i\omega_{nm}(t-\tau)}\sigma_{nm}\left(t-\tau\right)\approx e^{i\omega_{nm}t}\sigma_{nm}\left(t\right)$,\citep{nitzan2006chemical}
we arrive at:
\begin{align}
\frac{\partial\sigma_{nm}}{\partial t} & =-i\omega_{nm}\sigma_{nm}-\frac{i}{\hbar}\left[\sum_{l\neq n}\left\langle W_{nl}\right\rangle _{{\rm B}}\sigma_{lm}-\sum_{l\neq m}\sigma_{nl}\left\langle W_{lm}\right\rangle _{{\rm B}}\right]\nonumber \\
 & +\sum_{kl}\int_{0}^{t}d\tau\left[-M_{nk,kl}\left(\tau\right)e^{i\omega_{lk}\tau}\sigma_{lm}\left(t\right)-M_{kl,lm}\left(-\tau\right)e^{i\omega_{lk}\tau}\sigma_{nk}\left(t\right)\right.\nonumber \\
 & \left.+M_{nk,lm}\left(\tau\right)e^{i\omega_{kn}\tau}\sigma_{kl}\left(t\right)+M_{nk,lm}\left(-\tau\right)e^{i\omega_{ml}\tau}\sigma_{kl}\left(t\right)\right].
\end{align}
Here we could see that, the so-called Redfield tensor: $R_{nm,kl}\left(\omega,t\right)=\int_{0}^{t}d\tau M_{nm,kl}\left(\tau\right)e^{i\omega\tau}$
depends on time.

\section{Calculating $\left\langle W_{nm}\left(t\right)W_{kl}\left(0\right)\right\rangle _{\text{B}}$}
We represent the bath position ($q_{\alpha}$) and momentum ($p_{\alpha}$)
operators using raising and lowering operators ($b_{\alpha}^{\dagger}$
and $b_{\alpha}$, respectively):
\begin{align}
q_{\alpha} & =\sqrt{\frac{\hbar}{2\omega_{\alpha}}}\left(b_{\alpha}^{\dagger}+b_{\alpha}\right),\\
p_{\alpha} & =i\sqrt{\frac{\hbar\omega_{\alpha}}{2}}\left(b_{\alpha}^{\dagger}-b_{\alpha}\right).
\end{align}
Recall the dressed phonon coupling element, $W_{nm}$,
\begin{equation}
W_{nm}=\exp\left(-\frac{i}{\hbar}\sum_{\gamma}\frac{p_{\gamma}\tilde{V}_{nn}^{\gamma}}{\omega_{\gamma}^{2}}\right)\left[\sum_{\alpha}\tilde{V}_{nm}^{\alpha}q_{\alpha}\right]\exp\left(+\frac{i}{\hbar}\sum_{\xi}\frac{p_{\xi}\tilde{V}_{mm}^{\xi}}{\omega_{\xi}^{2}}\right).
\end{equation}
In the above equation, $\tilde{V}_{nm}^{\alpha}$, the coupling matrix
element between two polaritonic states $\left|\varphi_{n}\right\rangle $
and $\left|\varphi_{m}\right\rangle $ and phonon mode $\alpha$.
$\tilde{V}_{nm}^{\alpha}$ was obtained by applying the unitary transformation
$U$ that diagonalizes $H_{{\rm S}}$ to $\tilde{V}^{\alpha}=U^{\dagger}V^{\alpha}U$,
for each mode $\alpha$ (see main text). The corresponding correlation
function, $\left\langle W_{nm}\left(t\right)W_{kl}\left(0\right)\right\rangle _{\text{B}}$
(Eq.~(9)) can then be expressed as : 
\begin{align}
\left\langle W_{nm}\left(t\right)W_{kl}\left(0\right)\right\rangle _{\text{B}} & =\left\langle \left\{ \exp\left(-\sum_{\alpha}\frac{i\tilde{V}_{nn}^{\alpha}p_{\alpha}\left(t\right)}{\hbar\omega_{\alpha}^{2}}\right)\left[\sum_{\alpha}\tilde{V}_{nm}^{\alpha}q_{\alpha}\left(t\right)\right]\exp\left(\sum_{\alpha}\frac{i\tilde{V}_{mm}^{\alpha}p_{\alpha}\left(t\right)}{\hbar\omega_{\alpha}^{2}}\right)\right\} \right.\nonumber \\
 & \left.\left\{ \exp\left(-\sum_{\alpha}\frac{i\tilde{V}_{kk}^{\alpha}p_{\alpha}}{\hbar\omega_{\alpha}^{2}}\right)\left[\sum_{\alpha}\tilde{V}_{kl}^{\alpha}q_{\alpha}\right]\exp\left(\sum_{\alpha}\frac{i\tilde{V}_{ll}^{\alpha}p_{\alpha}}{\hbar\omega_{\alpha}^{2}}\right)\right\} \right\rangle _{\text{B}}\nonumber \\
 & =\left\langle e^{\Omega_{n}\left(t\right)}L^{\left(nm\right)}\left(t\right)e^{-\Omega_{m}\left(t\right)}e^{\Omega_{k}}L^{\left(kl\right)}e^{-\Omega_{l}}\right\rangle _{\text{B}},
\end{align}
where
\begin{align}
q_{\alpha}\left(t\right) & =\sqrt{\frac{\hbar}{2\omega_{\alpha}}}\left(b_{\alpha}^{\dagger}e^{i\omega_{\alpha}t}+b_{\alpha}e^{-i\omega_{\alpha}t}\right),\\
p_{\alpha}\left(t\right) & =i\sqrt{\frac{\hbar\omega_{\alpha}}{2}}\left(b_{\alpha}^{\dagger}e^{i\omega_{\alpha}t}-b_{\alpha}e^{-i\omega_{\alpha}t}\right),
\end{align}
so
\begin{align}
\Omega_{n}\left(t\right) & =\sum_{\alpha}\tilde{d}_{\alpha}^{\left(n\right)}\left(a_{\alpha}^{\dagger}e^{i\omega_{\alpha}t}-a_{\alpha}e^{-i\omega_{\alpha}t}\right),\\
L^{\left(nm\right)}\left(t\right) & =\sum_{\alpha}\tilde{c}_{\alpha}^{\left(nm\right)}\left(a_{\alpha}^{\dagger}e^{i\omega_{\alpha}t}+a_{\alpha}e^{-i\omega_{\alpha}t}\right),
\end{align}
where
\begin{align}
\tilde{d}_{\alpha}^{\left(n\right)} & =\frac{\tilde{V}_{nn}^{\alpha}}{\hbar\omega_{\alpha}^{2}}\sqrt{\frac{\hbar\omega_{\alpha}}{2}},\\
\tilde{c}_{\alpha}^{\left(nm\right)} & =\tilde{V}_{nm}^{\alpha}\sqrt{\frac{\hbar}{2\omega_{\alpha}}}.
\end{align}

Using the commutation relationship:
\begin{equation}
\left[\left(\alpha a_{j}^{\dagger}+\beta a_{j}\right),e^{\left(\gamma a_{j}^{\dagger}+\delta a_{j}\right)}\right]=\left[\left(\alpha a_{j}^{\dagger}+\beta a_{j}\right),\left(\gamma a_{j}^{\dagger}+\delta a_{j}\right)\right]e^{\left(\gamma a_{j}^{\dagger}+\delta a_{j}\right)},
\end{equation}
the correlation function can be expressed as
\begin{equation}
\left\langle W_{nm}\left(t\right)W_{kl}\left(0\right)\right\rangle _{\text{B}}=\left[h\left(t\right)+g\left(t\right)\right]f\left(t\right),
\end{equation}
where
\begin{align}
h\left(t\right) & =\sum_{\alpha}\left\{ -\left(\tilde{d}_{\alpha}^{(k)}-\tilde{d}_{\alpha}^{(l)}\right)\tilde{c}_{\alpha}^{(nm)}e^{i\omega_{\alpha}t}n_{\alpha}+\left(\tilde{d}_{\alpha}^{(k)}-\tilde{d}_{\alpha}^{(l)}\right)\tilde{c}_{\alpha}^{(nm)}e^{-i\omega_{\alpha}t}\left(n_{\alpha}+1\right)-\tilde{c}_{\alpha}^{(nm)}\left(\tilde{d}_{\alpha}^{(n)}+\tilde{d}_{\alpha}^{(m)}\right)\right\} \times\nonumber \\
 & \times\sum_{\alpha}\left\{ \tilde{c}_{\alpha}^{(kl)}\left(\tilde{d}_{\alpha}^{(n)}-\tilde{d}_{\alpha}^{(m)}\right)e^{i\omega_{\alpha}t}n_{\alpha}-\tilde{c}_{\alpha}^{(kl)}\left(\tilde{d}_{\alpha}^{(n)}-\tilde{d}_{\alpha}^{(m)}\right)e^{-i\omega_{\alpha}t}\left(n_{\alpha}+1\right)-\tilde{c}_{\alpha}^{(kl)}\left(\tilde{d}_{\alpha}^{(k)}+\tilde{d}_{\alpha}^{(l)}\right)\right\} ,
\end{align}
\begin{equation}
g\left(t\right)=\sum_{\alpha}\tilde{c}_{\alpha}^{(nm)}\tilde{c}_{\alpha}^{(kl)}\left[\left(n_{\alpha}+1\right)e^{-i\omega_{\alpha}t}+n_{\alpha}e^{i\omega_{\alpha}t}\right],
\end{equation}
\begin{align}
f\left(t\right) & =\prod_{\alpha}\exp\left\{ -\left[\left(\tilde{d}_{\alpha}^{(n)}-\tilde{d}_{\alpha}^{(m)}\right)^{2}+\left(\tilde{d}_{\alpha}^{(k)}-\tilde{d}_{\alpha}^{(l)}\right)^{2}+2\left(\tilde{d}_{\alpha}^{(n)}-\tilde{d}_{\alpha}^{(m)}\right)\left(\tilde{d}_{\alpha}^{(k)}-\tilde{d}_{\alpha}^{(l)}\right)\cos(\omega_{\alpha}t)\right]\left(n+1/2\right)\right.\nonumber \\
 & +\left.i\left(\tilde{d}_{\alpha}^{(n)}-\tilde{d}_{\alpha}^{(m)}\right)\left(\tilde{d}_{\alpha}^{(k)}-\tilde{d}_{\alpha}^{(l)}\right)\sin\left(\omega_{\alpha}t\right)\right\} ,
\end{align}
and $n=\left\langle a^{\dagger}a\right\rangle =\left(e^{\beta\hbar\omega}-1\right)^{-1}$.

Similarly, the average $\left\langle W_{nm}\right\rangle _{\text{B}}$
is given by: 
\begin{equation}
\left\langle W_{nm}\right\rangle _{\text{B}}=\left\langle e^{\Omega_{n}}L^{(nm)}e^{-\Omega_{m}}\right\rangle _{\text{B}}=\sum_{\alpha}\left(-\tilde{c}_{\alpha}^{(nm)}\tilde{d}_{\alpha}^{(n)}-\tilde{c}_{\alpha}^{(nm)}\tilde{d}_{\alpha}^{(m)}\right)\prod_{\alpha}\exp\left\{ -\left(\tilde{d}_{\alpha}^{(n)}-\tilde{d}_{\alpha}^{(m)}\right)^{2}\left(n_{\alpha}+\frac{1}{2}\right)\right\} .
\end{equation}

\section{Effective Coupling}
\subsection{Three level system}
We consider a three level system representing a donor, acceptor coupled
directly and both are coupled to a bridge state (cavity). The system
Hamiltonian is written as:
\begin{equation}
H=\left(\begin{array}{ccc}
E_{B} & g & g\\
g & E_{D} & J\\
g & J & E_{A}
\end{array}\right),
\end{equation}
where the donor and acceptor states with the energies $E_{D}$ and
$E_{A}$ are directly coupled by $J$. The bridge state couples to
the donor and acceptor with coupling strength $g$. The goal is to
rewrite the eigenvalue problem
\begin{equation}
\left[\left(\begin{array}{ccc}
E_{B} & g & g\\
g & E_{D} & J\\
g & J & E_{A}
\end{array}\right)-\boldsymbol{I}\lambda\right]\left(\begin{array}{c}
d_{B}\\
d_{D}\\
d_{A}
\end{array}\right)=0.
\end{equation}
in the donor-acceptor subspace. The three eigenvalues (we assume for
clarity that $E_{D}=E_{A}=E$) are given by:
\begin{align}
\lambda_{1} & =E-J,\\
\lambda_{2} & =\frac{1}{2}\left(E_{B}+E+J-\sqrt{8g^{2}+\left(E_{B}-E-J\right)^{2}}\right),\\
\lambda_{3} & =\frac{1}{2}\left(E_{B}+E+J+\sqrt{8g^{2}+\left(E_{B}-E-J\right)^{2}}\right).
\end{align}
The corresponding eigenvectors can be obtained from:
\begin{align}
\left(E_{B}-\lambda\right)d_{B}+gd_{D}+gd_{A} & =0,\\
gd_{B}+\left(E-\lambda\right)d_{D}+Jd_{A} & =0,\\
gd_{B}+Jd_{D}+\left(E-\lambda\right)d_{A} & =0.
\end{align}
Replacing $d_{B}$ in the last two equations by the solution of the
first one, $d_{B}=\frac{gd_{D}+gd_{A}}{\left(\lambda-E_{B}\right)}$,
we find:
\begin{align}
\left[\frac{g^{2}}{\left(\lambda-E_{B}\right)}+E-\lambda\right]d_{D}+\left[\frac{g^{2}}{\left(\lambda-E_{B}\right)}+J\right]d_{A} & =0,\\
\left[\frac{g^{2}}{\left(\lambda-E_{B}\right)}+J\right]d_{D}+\left[\frac{g^{2}}{\left(\lambda-E_{B}\right)}+E-\lambda\right]d_{A} & =0.
\end{align}
which can be rewritten in a similar form to an eigenvalue problem
in the Hilbert space of the donor and acceptor:
\begin{equation}
\left(\begin{array}{cc}
E_{D}^{\text{eff}}-\lambda & J_{\text{eff}}\\
J_{\text{eff}} & E_{A}^{\text{eff}}-\lambda
\end{array}\right)\left(\begin{array}{c}
d_{D}\\
d_{A}
\end{array}\right)=0,
\end{equation}
where
\begin{align}
E_{D}^{\text{eff}} & =E_{A}^{\text{eff}}=\frac{g^{2}}{\left(\lambda-E_{B}\right)}+E,
\end{align}
and the effective coupling
\begin{equation}
J_{\text{eff}}=\frac{g^{2}}{\left(\lambda-E_{B}\right)}+J
\end{equation}
depend on $\lambda.$ For the two eigenvalues $\lambda_{2}$ and $\lambda_{3}$we
find:
\begin{align}
J_{\text{eff}}^{(2,3)} & =J+\frac{2g^{2}}{\left(E+J-E_{B}\right)\left(1\mp\sqrt{1+2\left(\frac{2g}{E+J-E_{B}}\right)^{2}}\right)}.
\end{align}
For small values $g\ll J$, $J_{\text{eff}}^{(2,3)}\rightarrow J$
and for large values of $g\gg J$, $J_{\text{eff}}^{(2,3)}\rightarrow\mp\frac{g}{\left(\sqrt{2}\right)}$.
This implies that for large coupling of the donor and acceptor to
the bridge results in an effective direct coupling between the donor
and acceptor that scales with $g$.

\bibliography{reference_SI}